\documentstyle[psfig,rotate]{l-aa}
\newcommand{\nodata}{\multicolumn{1}{c}{$\cdots$}}
\newcommand{\philband}{$\rm A^{1}\Pi_{u}-X^{1}\Sigma^{+}_{g}$}
\newcommand{\swanband}{$\rm d^{3}\Pi_{g}-a^{3}\Pi_{u}$ }
\newcommand{\redband}{$\rm A^{2}\Pi-X^{2}\Sigma^{+}$}
\newcommand{\sysch}{$\rm A^{1}\Pi-X^{1}\Sigma^{+}$}

\begin{document}
\thesaurus{07(08.16.4; 08.03.4; 02.12.1; 02.13.5)}
\title{Circumstellar C$_{2}$, CN, and CH$^{+}$ in the optical spectra of 
post-AGB stars
\thanks{Based on observations with the Utrecht Echelle
                 Spectrograph on the William Herschel Telescope 
                 (La Palma, Spain), and the McDonald observatory 2.7m
                 telescope (Texas).}
\thanks{Tables 5 and A.1/2/3/4/5 are available in electronic form at the 
CDS via anonymous
ftp to cdsarc.u-strasbg.fr (130.79.128.5) or 
via http://cdsweb.u-strasbg.fr/Abstract.html, or
from the authors.}}

\author{Eric J. Bakker\inst{1,2,3}\and
        Ewine F. van Dishoeck\inst{4}\and
        L.B.F.M. Waters\inst{5,6}\and
        Ton Schoenmaker\inst{7}}

\offprints{Eric J. Bakker at the University of Texas, 
           ebakker@viking.as.utexas.edu}

\institute{University of Texas,
           Department of Astronomy,
           TX78712, USA
           \and
           Astronomical Institute, University of Utrecht,
           P.O. Box 80000,
           NL-3508 TA Utrecht,
           The Netherlands
           \and
           SRON Laboratory for Space Research,
           Sorbonnelaan 2,
           NL-3584 CA Utrecht,
           The Netherlands 
           \and
           Sterrewacht Leiden, 
           University of Leiden,
           P.O. Box 9513,
           NL-2300 RA Leiden,
           The Netherlands
           \and
           Astronomical Institute,
           University of Amsterdam,
           Kruislaan 403,
           NL-1098 SJ Amsterdam,
           The Netherlands
           \and 
           SRON Laboratory for Space Research,
           P.O. Box 800,
           NL-9700 AV Groningen,
           The Netherlands  
           \and
           Kapteyn Sterrenwacht Roden,
           Mensingheweg 20,
           NL-9301 KA Roden,
           The Netherlands}

\date{Received 16 April 1996/ October 2 1996}
\maketitle

 \begin{abstract}
 We present optical high-resolution spectra 
 of a sample of sixteen post-AGB stars and IRC~+10216. Of the post-AGB
 stars, ten
 show C$_{2}$ Phillips 
 (\philband) and Swan (\swanband)
 and CN Red System (\redband) absorption, one CH$^{+}$ (\sysch) 
 emission, one CH$^{+}$ absorption, and four without any molecules.
 We find typically $T_{\rm rot} \sim 43-399,  155-202$, and $18-50 $~K, 
 $\log N \sim
 14.90-15.57, 14.35$, and $15.03-16.47$~cm$^{-2}$
 for C$_{2}$, CH$^{+}$, and CN respectively,
 and $0.6 \leq N$(CN)/$N$(C$_{2}) \leq 11.2$. We did not detect isotopic
 lines, which places a lower limit on the isotope ratio of 
 ${\rm ^{12}C/^{13}C}>20$.
 The presence of C$_{2}$ and CN absorption is correlated with cold dust 
 ($T_{\rm dust}\leq 300$~K) and the presence of CH$^{+}$ 
 with  hot dust ($T_{\rm dust}\geq 300$~K).
 All objects  with the unidentified 21~$\mu$m emission   feature 
 exhibit C$_{2}$ and CN absorption, but not all objects with C$_{2}$ and 
 CN detections exhibit a 21~$\mu$m feature. 
 The derived expansion velocity, ranging from $5$ to $44$ km s$^{-1}$,
 is the same as that derived from CO
 millimeter line emission. This unambiguously proves that these lines 
 are of circumstellar origin and are formed in the AGB ejecta
 (circumstellar shell expelled during the preceding AGB phase).
 Furthermore there seems to be   a relation between the C$_{2}$ 
 molecular column density and the expansion velocity,
 which is attributed to the fact that a higher  
 carbon abundance of the dust leads to a more efficient
 acceleration of the AGB wind.
 Using simple assumptions for the location of the molecular lines 
 and molecular abundances, mass-loss rates have been derived from the
 molecular absorption lines and are 
 comparable to those obtained from CO emission lines and the infrared excess.

\end{abstract}

\section{Introduction}

The first study  of molecules in the optical spectrum of a post-AGB
star concerned the presence of 
C$_{3}$ absorption and C$_{2}$ emission in the reflected light
of the lobes of the Cygnus Egg Nebula (Crampton et al.  1975).
Renewed interest was triggered by the discovery 
by Waelkens et al.  (1992), Balm \& Jura (1992), and Hall et al. (1992)
of CH$^{+}$ emission in the optical spectrum of the famous Red Rectangle.
Recently, Waelkens et al.  (1995) reported
on the presence of CH$^{+}$  absorption  in HD~213985.
In this paper (which we will
refer to as Paper~II) we will study the presence of C$_{2}$, CN, and
CH$^{+}$ in the optical spectrum of thirteen (post-)AGB stars.

Bakker et al.  (1996c, Paper~I) were the first to analyze the molecular bands 
of C$_{2}$ and CN in the spectrum of the post-AGB star 
HD~56126 and showed that the expansion 
velocity and the excitation conditions are consistent with the lines being 
formed in the AGB ejecta. Extending this work, Bakker et al. (1995) showed that
the same is valid for the molecular absorption lines  in four other post-AGB 
stars (IRAS~04296+3429, IRAS~05113+1347, IRAS~08005-2356, and AFGL~2688)
and argued on the basis of the relation
between CO (or OH for IRAS~08005-2356) and C$_{2}$ or 
CN expansion velocity that these molecular absorption
lines are formed in 
the AGB ejecta and are therefore of circumstellar origin.
Hrivnak (1995)  discussed the presence of C$_{2}$  and C$_{3}$ 
in the low-resolution spectra of nine post-AGB stars. 

Theoretical models of the formation and dissociation of molecules in the
extended envelope of the carbon-rich AGB star
IRC~+10216 by Cherchneff et al.  (1993)  showed
that, e.g., C$_{2}$ and CN are only present in a thin shell
of material within the extended envelope. 
Close to the star carbon and nitrogen are locked up in
complex stable molecules such as C$_2$H$_2$ and HCN, while at larger
distances the 
interstellar ultraviolet radiation field photodissociates 
these molecules to C$_{2}$ and CN. At even larger distances
the UV radiation field photodissociates simple molecules into their
constituent atoms and ions. The net effect is that simple molecules 
exist only  in a thin shell of material.

Interstellar C$_{2}$ has been discussed extensively by van Dishoeck \&
Black (1982). They have shown that the 
excitation of C$_{2}$ is a sensitive balance between photoexcitation and 
collisional (de-)excitation. By 
modeling the excitation, 
the relative population over the rotational energy levels of the C$_{2}$
ground state can be used to  determine 
the particle density, radiation field, and kinetic temperature
of the line-forming region.  In a separate paper (Paper~III in preparation) we will
apply this model to the stars studied, while  an earlier account of this 
work can be found  in Bakker et al.  (1995).

In \S~2 we discuss the criteria used in
selecting our sample of mainly carbon-rich post-AGB stars, and describe
the observations and data reduction. 
\S~3 describes the method used in 
identifying molecular bands and determination of 
the expansion velocities of the AGB ejecta.  
Rotational diagrams are used to derive molecular rotational temperatures,
column densities,  and mass-loss rates. 
The results are discussed (\S~4) for the sample as a whole and
for each star separately. We will finish 
with a short conclusion (\S~5).

\begin{table*} % tab 1
\caption{Infrared fluxes and spectral features of the seventeen program stars.} 
\begin{tabular}{llrrrllll}
\hline
\hline
Object         &Sp.T. &   $m_{v}$&$f_{12}$ [Jy]&$f_{25}$ [Jy]      
                                                    &3.3 $^\ast$
                                                       &3.4-3.5 $^\ast$
                                                          &21 $^\ast$
                                                             &Remark         \\
\hline
               &      &     &           &           &  &  &  &               \\
IRAS~04296+3429&G0~Ia &14.21&   12.74   &   45.94   &e &e &e &\nodata        \\
IRAS~05113+1347&G8~Ia &14.40&    3.78   &   15.30   &e &nd&e &\nodata        \\
IRAS~05341+0852&F4~I  &12.8 &    4.51   &   9.85    &e &e &e &s-process enhanced \\
HD~44179       &B9~I  & 8.84&  421.60   &  456.10   &e &nd&nd&IRAS~06176-1036; 
                                                               Red Rectangle \\
HD~52961       &F8~I  & 8.50&    4.53   &    2.22   &no&no&nd&IRAS~07008+1050\\
HD~56126       &F5~I  & 8.23&   24.51   &  116.70   &e &nd&e &IRAS~07134+1005\\
IRAS~08005-2356&F5~I  &11.46&   17.96   &   51.80   &no&no&nd&OH maser       \\
IRC~+10216     &C9.5  &11.00&47530.00   &23070.00   &no&no&nd&IRAS~09452+1330 
                                                                (carbon star)\\
HR~4049        &F1~I  & 5.52&   48.25   &    9.55   &e &nd&nd&IRAS~10158-2844\\
HD~161796      &F3~Ib & 7.27&    6.12   &  183.50   &no&no&nd&IRAS~17436+5003
                                                                (oxygen rich)\\
IRAS~20000+3239&G8~Ia &13.40&   15.03   &   70.97   &no&no&e &\nodata        \\
AFGL~2688      &F5~Iae&14.00&  339.00   & 3041.00   &e &no&e &Egg Nebulae    \\
IRAS~22223+4327&G0~Ia &13.30&    2.12   &   37.10   &no&e &e &\nodata        \\
HD~235858      &G2~Ia & 9.30&   73.88   &  302.40   &e &e &e &IRAS~22272+5435\\
HD~213985      &B9~I  & 8.83&    5.57   &    4.66   &te&nd&nd&IRAS~22327-1731\\
BD~+39$^o$4926 &F1~I  & 9.24&      nd   &      nd   &no&no&no&\nodata        \\
IRAS~23304+6147&G2~Ia &13.15&   11.36   &   59.07   &no&no&e &\nodata        \\
               &      &     &           &           &  &  &  &               \\
\hline
\hline
\multicolumn{9}{l}{$^\ast$: e: emission; nd: not detected; no: not observed,
3.3, 3.4-3.5, and 21 $\mu$m infrared features} \\
\end{tabular}
\end{table*}

\section{Observations and data reduction}

\subsection{Selection of program stars}

After the serendipitous discovery of C$_{2}$ and CN in the optical
spectrum of HD~56126 (Paper~I) we 
started an extensive observing campaign in order to find additional
post-AGB stars with molecular absorption or emission lines. 
Assuming that the 
presence of carbon-based molecules is related to the high carbon abundance
of the AGB ejecta, we have selected those
stars with a carbon-rich circumstellar environment. The  presence
of the 3.3 and 3.4-3.5~$\mu$m Polycyclic Aromatic Hydrocarbon (PAH) features
and of the unidentified 21~$\mu$m feature 
(Kwok et al.  1989) were used as criteria for the carbon-rich 
nature of the AGB ejecta (Table~1). Recently the number of objects exhibiting
the 21 $\mu$m feature has been extended (e.g., Henning et al. 1996,  Justtanont et al. 1996),
and these objects have been selected for a follow-up study.
The sample was supplemented with those stars of 
special importance for the theory of post-AGB evolution (e.g., the metal-depleted
post-AGB binaries: HD~52961, HR~4049, BD~+39$^{o}$4926, Red Rectangle, and HD~213985).
Two O-rich post-AGB stars (e.g,  HD~161796 and IRAS~08005-2356)
were added to see whether O-rich star have the carbon based molecules, and
the well studied carbon star IRC+10216
was added to the list, since its spectrum could possibly be used 
as a template for identifying molecular features. 
With a limiting magnitude of $m_{v}=14$  this resulted in a list 
of sixteen Post-AGB stars (and IRC~+10216) observable from La Palma. 
IRAS~05341+0852 was added to the list at a later  stage. The optical spectrum
and the molecular bands are described by Reddy et al. (1997). 

The observed stars are listed in Table~1, and 
the resulting spectra in three selected windows 
are presented in Figs.~1,2, and 3.
To facilitate comparison  {\bf all} velocities given
are heliocentric (marked with the symbol $v_{\odot}$)
and the corrections were obtained
with the Starlink RV utility Version 2.2.

\subsection{Observations and data reduction}

\subsubsection{WHT/UES}

The observations were made in four different runs between February
1992 and August 1994 with the Utrecht Echelle Spectrograph (UES)
mounted on the Nasmyth platform of the 4.2m William Herschel
Telescope (WHT) on La Palma. The echellograms of February 1992 were
recorded on an EEV CCD-05-30 detector with $1242\times1152$ pixels of
$22.5\times22.5$ $\mu$m$^{2}$ each, 
while in 1994 a Tektronix TK1024A device was
used with $1024\times1024$ pixels of $24\times24$ $\mu$m$^2$ each. 
The wavelength
coverage of the UES ranges from $3000$ to $11000$~\AA, and even
using the echelle with 31.6 lines per mm, one needs several settings
because of the small frame size of the CCD's. For our observations we used
settings with central wavelengths of 4020, 5261 and 7127 \AA,
giving a wavelength coverage from about 3650 to 10000 \AA, with
some overlap. For each setting Tungsten flatfield and 
Thorium-Argon (Th-Ar) calibration exposures were taken.

The echellograms from the UES were reduced by TS using the echelle
package of IRAF V2.9 running on a DEC3100 workstation. All frames
were bias corrected using the overscan columns of the CCD's.
The traced orders were extracted with the optimal extraction method
using a variance weighting algorithm. Subsequently, the extracted
spectra were divided by the identically extracted flatfields and
normalized to the continuum.
The wavelength calibration was performed using the FIGARO list of
Th-Ar lines. About 700 calibration lines, evenly distributed
over the orders, were used in a two-dimensional fit with typical rms
residuals of 0.25 km s$^{-1}$ (about 0.1 pixel). By checking the wavelength
of a number of telluric oxygen absorption lines, the drift between
calibration and target frames was found to be less than 1 km s$^{-1}$.

The resolution  was determined from a number of identified telluric lines
(Moore et al.  1966) yielding a $FWHM$ of the line profile of
$6.0\pm1.0$~km s$^{-1}$ and a spectral resolving power  of $R \sim 5 \times 10^{4}$. 
The typical signal-to-noise ratio is $SNR=100$, and the minimum equivalent 
width detectable is about 7~m\AA~ (see App.~A only at CDS).

\subsubsection{McDonald/CS21}

In search for the CH$^{+}$ band towards HD~56126 (no such band was detected),
we have obtained high-resolution spectra using the 
2DCOUD\'{E} (CS21) spectrograph (Tull et al. 1995)
of the 2.7m telescope at the McDonald observatory. Light was fed to the TK3
CCD, $2048\times 2048$ pixels with each  $24 \times 24$ $\mu$m$^{2}$,  at the
F1 focus using E2 echelle  having 52.6759 grooves~mm$^{-1}$. 
The spectra were reduced by EJB using the data reduction package
IRAF: bias and scattered light subtracted, 
flat fielded and wavelength calibrated
using a ThAr arc spectrum. The wavelength calibration has a rms internal error
of 0.42 m\AA~ which corresponds to 0.016~km~s$^{-1}$ at the 
wavelength of interest. The resolution was determined from the Thorium lines 
as $R \sim 160,000$.

\begin{figure*}  % Fig. 1
\centerline{\hbox{\psfig{figure=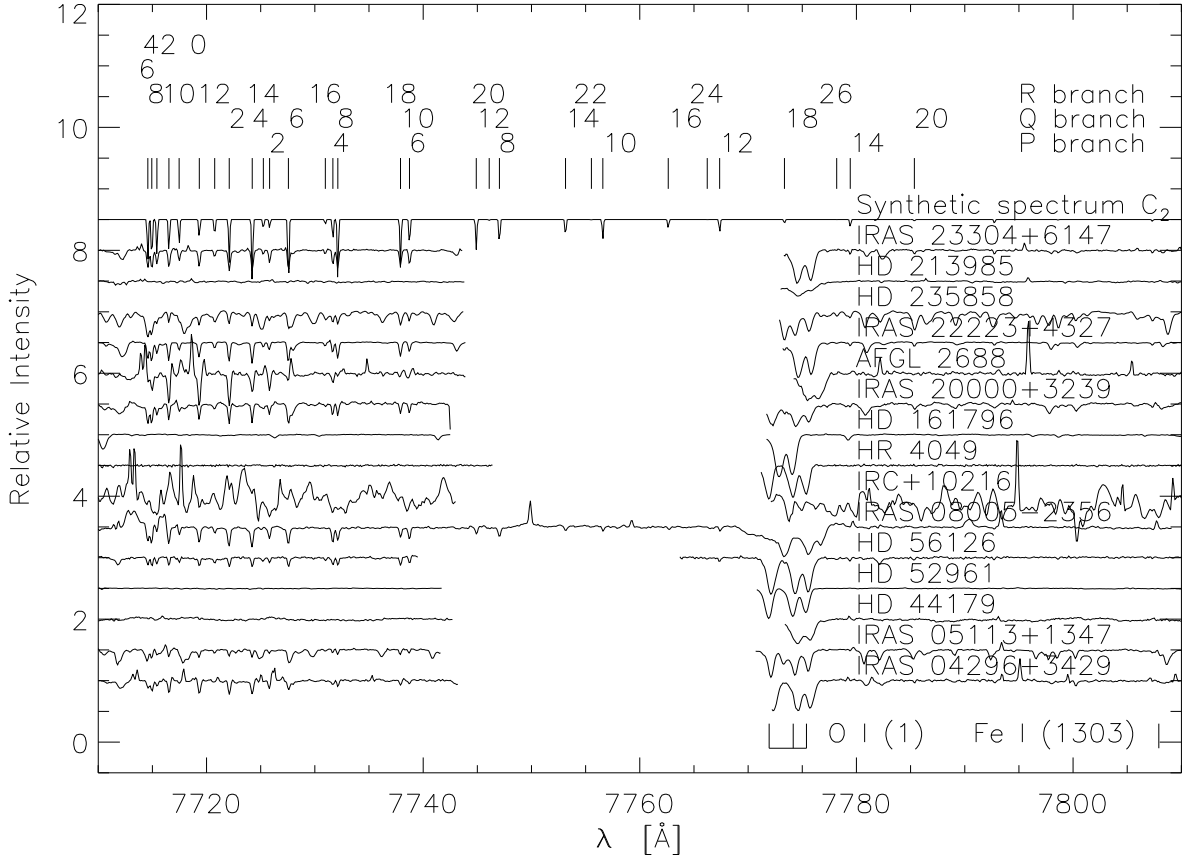,width=\textwidth}}}
\caption{Spectra of the observed stars in the wavelength region of 
the C$_{2}$ $\rm A^{1}\Pi_{u}-X^{1}\Sigma^{+}_{g}$
(3,0) band. The top spectrum is a synthetic spectrum computed
using $T_{\rm rot}=200$~K, $\log N = 15.60$~cm$^{-2}$
and $R\sim5\times10^{4}$. All other spectra are corrected
for the velocity of the molecular lines such that the molecular
lines are at their rest wavelengths. The three absorption lines near
7775~\AA~ are due to the OI(1) multiplet. Note that the wavelength part from
7740~\AA~ to 7770~\AA~ is not observed for most of the objects in the
sample.}
\end{figure*}

\begin{figure*} % Fig. 2
\centerline{\hbox{\psfig{figure=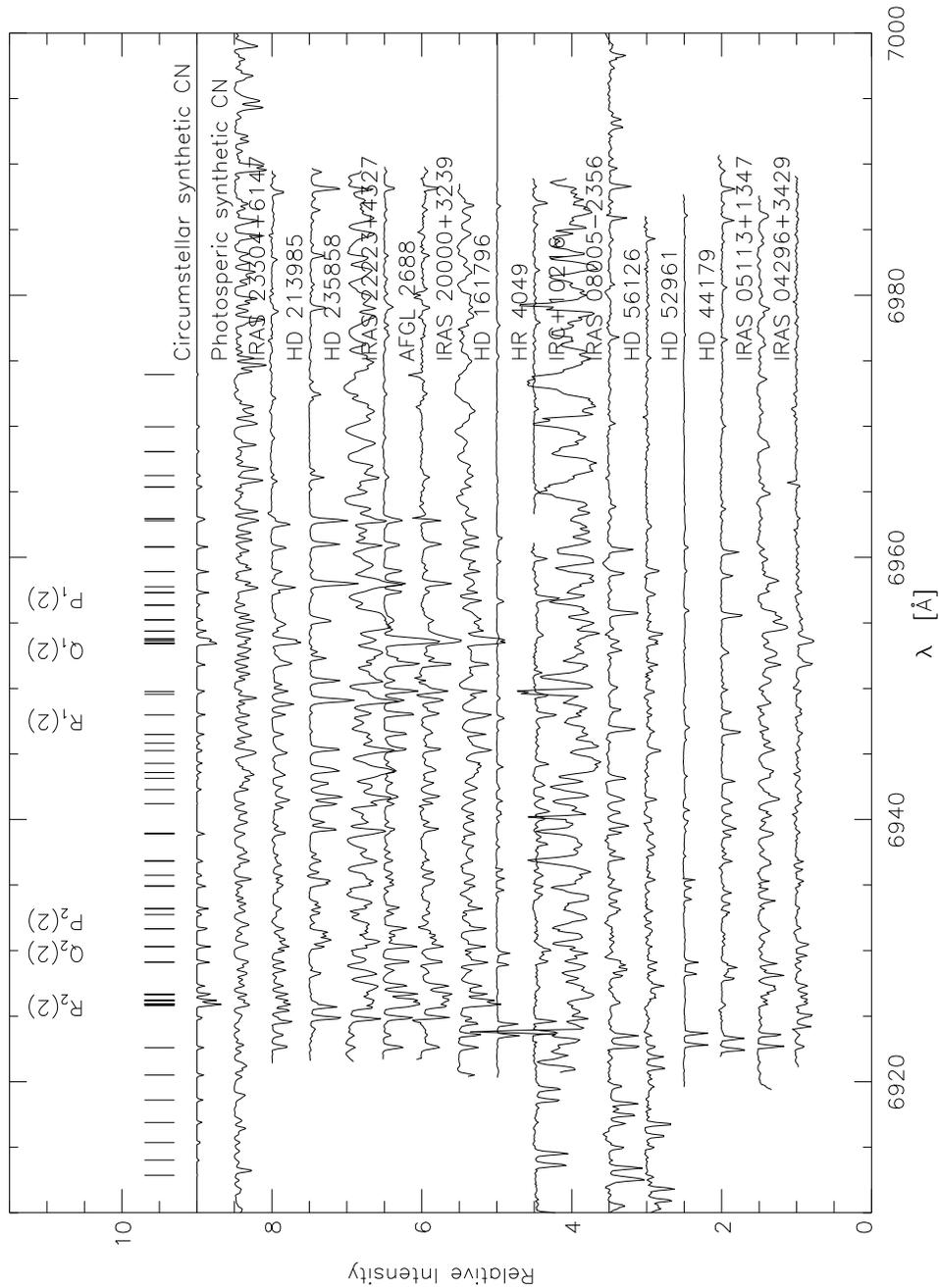,height=20cm,angle=180}}}
\caption{Spectra of the observed stars in the wavelength region of 
the CN $\rm A^{2}\Pi-X^{2}\Sigma^{+}$
(3,0) band. The top spectrum is a synthetic spectrum of circumstellar
$^{12}$CN computed
using $T_{\rm rot}=25$~K, $\log N = 16.50$~cm$^{-2}$
and $R\sim5\times10^{4}$, followed by a synthetic spectrum
of photospheric $^{12}$CN 
using $T_{\rm rot}=5500$~K and $\log N = 16.78$~cm$^{-2}$.
All spectra are corrected
for the velocity of the molecular lines such that the 
circumstellar molecular
lines are at their rest wavelengths. Most of the unidentified 
narrow absorption lines (see HR~4049 which has no circumstellar
or photospheric lines in this part of the spectrum)
are due to telluric O$_{2}$ and H$_{2}$O (Moore et al. 1966)
or photospheric CN (only for HD~235858,  IRAS~20000+3239,
and IRAS~05113+13477).}
\end{figure*}

\begin{figure*} % Fig. 3
\centerline{\hbox{\psfig{figure=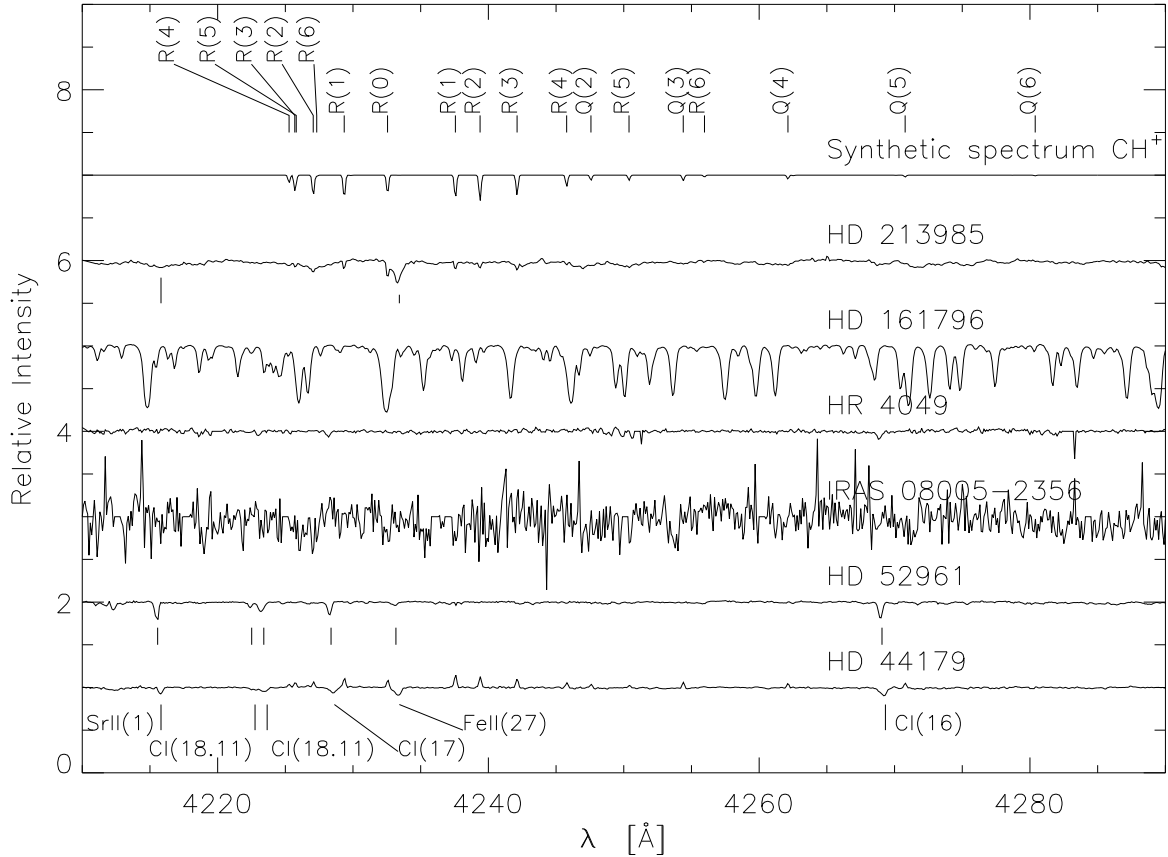,width=\textwidth}}}
\caption{Spectra of the observed stars in the wavelength region of 
the CH$^{+}$  $\rm A^{1}\Pi-X^{1}\Sigma^{+}$ 
(0,0) band. The top spectrum is a synthetic spectrum computed
using $T_{\rm rot}=200$~K, $\log N =14.70$~cm$^{-2}$
and $R=5\times10^{4}$. All other spectra are corrected
for the velocity of the molecular lines such that the molecular
lines are at their rest wavelengths. Note the broad photospheric absorption
features in the spectrum of HD~213985 (e.g., SrII(1) and the 
somewhat narrower FeII(27) feature.)}
\end{figure*}

\begin{table} % tab 2
\caption{Circumstellar molecular bands in the optical spectra of post-AGB stars.}
\begin{tabular}{rllllr}
\hline
\hline
Molecule             &System     &Band            &($v',v''$)
                                                            &\multicolumn{2}{c}{$\lambda$ range}\\
                     &           &                &         &\multicolumn{2}{c}{[\AA]} \\
\hline 
                     &           &                &         &     &          \\
$^{12}$C$^{12}$C     &Phillips   &\philband       &(1,0)    &10137&10213     \\
$^{12}$C$^{13}$C     &Phillips   &\philband       &(1,0)    &10166&10242     \\
$^{12}$C$^{12}$C     &Phillips   &\philband       &(2,0)    & 8753& 8913     \\
$^{12}$C$^{13}$C     &Phillips   &\philband       &(2,0)    & 8797& 8957     \\
$^{12}$C$^{12}$C     &Phillips   &\philband       &(3,0)    & 7717& 7809     \\
$^{12}$C$^{13}$C     &Phillips   &\philband       &(3,0)    & 7769& 7861     \\
$^{12}$C$^{12}$C     &Phillips   &\philband       &(4,0)    & 7910& 7002     \\
$^{12}$C$^{13}$C     &Phillips   &\philband       &(4,0)    & 7965& 7057     \\
                     &           &                &         &     &          \\
$^{12}$C$^{12}$C     &Swan       &\swanband       &(0,0)    & 5100& 5166     \\
$^{12}$C$^{13}$C     &Swan       &\swanband       &(0,0)    & 5995& 5161     \\
$^{12}$C$^{12}$C     &Swan       &\swanband       &(1,0)    & 4710& 4740     \\
$^{12}$C$^{13}$C     &Swan       &\swanband       &(1,0)    & 4705& 4735     \\
                     &           &                &         &     &          \\
$^{12}$C$^{14}$N     &Red System &\redband        &(1,0)    & 9130& 9220     \\
$^{13}$C$^{14}$N     &Red System &\redband        &(1,0)    & 9159& 9249     \\
$^{12}$C$^{14}$N     &Red System &\redband        &(2,0)    & 7874& 7918     \\
$^{13}$C$^{14}$N     &Red System &\redband        &(2,0)    & 7918& 7962     \\
$^{12}$C$^{14}$N     &Red System &\redband        &(3,0)    & 6927& 6954     \\
$^{13}$C$^{14}$N     &Red System &\redband        &(3,0)    & 6967& 6994     \\
$^{12}$C$^{14}$N     &Red System &\redband        &(4,0)    & 6190& 6225     \\
$^{13}$C$^{14}$N     &Red System &\redband        &(4,0)    & 6244& 6279     \\
                     &           &                &         &     &          \\
$^{12}$C$^{1}$H$^{+}$&           &\sysch          &(0,0)    & 4220& 4280     \\
$^{13}$C$^{1}$H$^{+}$&           &\sysch          &(0,0)    & 4220& 4280     \\
$^{12}$C$^{1}$H$^{+}$&           &\sysch          &(1,0)    & 3950& 3990     \\
$^{13}$C$^{1}$H$^{+}$&           &\sysch          &(1,0)    & 3950& 3990     \\
                     &           &                &         &     &          \\
\hline
\hline
\end{tabular}
\end{table}

\section{Molecular absorption and emission lines}

\subsection{Identification of molecular bands}

Molecular absorption line profiles resemble closely  those of
telluric absorption lines, while molecular  emission line profiles
can easily be  confused with cosmic spikes. To prevent confusion
with cosmic spikes most integrations were divided into two 
or more shorter exposures, 
and the individual spectra were combined after the reduction process
using a median filter to remove cosmic spikes.
A  number of (hot) stars have been observed to identify telluric lines.

\begin{table*} % tab 3
\caption{The detection of molecular bands in the optical spectra 
of post-AGB stars.}
\begin{tabular}{llllllllllll}
\hline
\hline
Object&\multicolumn{2}{c}{C$_{2}$ Swan $^\ast$}
      &\multicolumn{3}{c}{C$_{2}$ Phillips$^\ast$ }
      &\multicolumn{4}{c}{CN Red System $^\ast$}
      &\multicolumn{2}{c}{CH$^{+}$$^\ast$ } \\
      &\multicolumn{2}{c}{\swanband}
      &\multicolumn{3}{c}{\philband}
      &\multicolumn{4}{c}{\redband} 
      &\multicolumn{2}{c}{\sysch} \\
      &(0,0)&(1,0)              
      &(1,0)&(2,0)&(3,0)        
      &(1,0)&(2,0)&(3,0)&(4,0) 
      &(0,0)&(1,0)              \\
\hline
               &  &  &  &  &  &  &  &  &  &  &   \\
IRAS~04296+3429&no&no&a &a &a &a &a &a &b &no&no \\
IRAS~05113+1347&a &b &a &a &a &a &a &a &b &no&no \\
IRAS~05341+0852&no&no&a &a &a &a &a &a &no&no&no \\
HD~44179       &np&np&np&np&np&np&np&np&np&e &e  \\
HD~52961       &np&np&np&np&np&np&np&np&np&np&np \\
HD~56126       &a &a &a &a &a &a &a &a &a &np&np \\
IRAS~08005-2356&ta&no&a &a &a &a &a &np &np&b &b \\
IRC~+10216     &b &b &a &a &a &a &ta&b &b &no&no \\
HR~4049        &np&np&np&np&np&np&np&np&np&np&np \\
HD~161796      &no&no&np&np&np&np&np&np&np&np&np \\
IRAS~20000+3239&a &b &no&a &a &a &a &a &a &no&no \\
AFGL~2688      &a &b &e &a &a &a &a &a &a &no&no \\
IRAS~22223+4327&a &a &no&a &a &a &a &a &a &no&no \\
HD~235858      &a &a &a &a &a &a &a &b &b &no&no \\
HD~213985      &np&np&np&np&np&np&np&np&np&a &a  \\
BD~+39$^o$4926 &no&no&no&no&no&no&no&no&no&np&no \\
IRAS~23304+6147&a &a &no&a &a &a &a &a &a &no&no \\
               &  &  &  &  &  &  &  &  &  &  &   \\
\hline
\hline
\multicolumn{12}{l}{$^\ast$ a: absorption; b: blended; e: emission;  no: not observed; 
np: not present; ta: tentative absorption} \\
\multicolumn{12}{l}{C is $^{12}$C, N is $^{14}$N, and H is $^{1}$H} \\
\end{tabular}
\end{table*}

In this paper we have restricted ourselves to twelve molecular bands 
and their $^{13}$C isotopic species
(Table~2) of three different species (C$_{2}$, CN, and CH$^{+}$).
Table~3 gives for each star observed and for each molecular band,
whether it was detected (in absorption or emission),
tentatively detected (mostly when the lines are severely blended
by photospheric lines), or not detected at all.
We have carefully studied all spectra  to find 
additional ``prominent'' molecular bands, but none were detected.
Of course this does not exclude the presence of weak molecular 
bands with equivalent widths of less than 10~m\AA.
IRAS~05113+1347, IRAS~20000+3239, and HD~235858 show also CN Red System band
photospheric absorption (Fig.~2). These lines are easily distinguisable
from the circumstellar CN lines, since the lines are much broader
and they are identified with much higher energy levels 
(typically $30 \leq N'' \leq 90$).

We have carefully looked for the isotopic lines, e.g., $^{12}$C$^{13}$C,
$^{13}$C$^{14}$N, and $^{13}$CH$^+$, with negative results. From this we 
deduce a typical lower limit on the isotope ratio of $^{12}$C/$^{13}$C $>20$.

In Paper~I we pointed out that optical depth effects play an important role for 
the stronger molecular bands; therefore in this study we have limited  the analysis to 
the weaker molecular bands: C$_{2}$ Phillips (3,0) (Fig.~1), CN Red System
(3,0) (Fig.~2), and CH$^{+}$ (0,0) (Fig.~3).

The wavenumbers of the   C$_{2}$ Phillips bands
have been taken from  Chauville et al. (1977) ((1,0) and (2,0)) and
Ballik \& Ramsay (1963) (3,0) and CH$^{+}$ (0,0) and (1,0) bands  from
Carrington \& Ramsay (1982).
Conversion from wavenumber in vacuum to wavelength in air was made by
by applying  Cauchy's formula.
Wavelengths of the CN Red System were extracted from the SCAN tape
(J{\o}rgenson \& Larsson 1990).

\subsection{Determination of the expansion velocity, rotational
temperature, column density, time-scale, and mass-loss rate}

\subsubsection{Expansion velocity}

The first step in analyzing a molecular band is to check that all 
the candidate molecular 
lines yield the same radial velocity. The average radial velocity of all 
lines is a good measure for the true velocity of the line-forming region. For 
absorption lines this is the line-of-sight to the star, while for  emission 
lines it can be anywhere within the ``slit'' (defined 
as the area on the CCD used for extracting the spectrum)
of the telescope.

\begin{table*} % tab 4
\caption{Adopted system and AGB outflow velocities of programs stars.}
\begin{tabular}{lrrrll}
\hline
\hline
Object&$v_{\rm sys,lsr}$ $\rm ^a$&$v_{\rm sys,\odot}$ $\rm ^b$
&$v_{\rm exp}$ $\rm ^c$&Remark &Reference\\
\hline
               &        &       &       &                               & \\
IRAS~04296+3429&$-66.0 $&$-59.0$&$12.0 $&CO($J=2-1$)                    &O\\
IRAS~05113+1347&$-12.0 $&$  2.0$&\nodata&photospheric lines             &B\\
IRAS~05341+0852&$  9.6 $&$ 25.0$&\nodata&photospheric lines             &R\\
HD~44179       &$  2.7 $&$ 20.7$& 6.0   &binary $P=310\pm3$~days        &V,J\\
HD~52961       &\nodata&\nodata &\nodata&pulsator $P=70.8$~days         &F\\
HD~56126       &$ 71.0 $&$ 85.6$&$10.0 $&CO                             &Z\\
IRAS~08005-2356&$ 46.7 $&$ 61.1$&$49.8 $&OH maser                       &T\\
IRC~+10216     &$-26.2 $&$-19.1$&$14.1 $&pulsator $P=638$~days          &H,D\\
HR~4049        &$-44.5 $&$-32.9$&\nodata&binary $P=429$~days            &V\\
HD~161796      &$-36.0 $&$ -0.4$&$11.5 $&CO($J=1-0$)                    &L\\
IRAS~20000+3239&$ 14.0 $&$ -4.1$&$12   $&CO($J=1-0$)                    &O\\
AFGL~2688      &$-33.3 $&$-49.2$&$22.8 $&\nodata                        &Y\\
IRAS~22223+4327&$-30.0 $&$-42.2$&$14.0 $&CO($1-0$)                      &O\\
HD~235858      &$-30.9 $&$-43.1$&$11.6 $&CO                             &Z\\
HD~213985      &$-42.5 $&$-45.7$&\nodata&binary $P=259$~days            &V\\
BD~+39$^o$4926 &\nodata &\nodata&\nodata&binary $P=775$~days, no infrared excess& K \\
IRAS~23304+6147&$-15.9 $&$-25.8$&$15.5 $&CO($2-1$)                      &W\\
               &        &       &       &                               & \\
\hline
\hline
\multicolumn{6}{l}{$\rm ^a$ $v_{\rm lsr}$ is the system velocity in the local 
                    standard of rest system [km~s$^{-1}$]} \\
\multicolumn{6}{l}{$\rm ^b$ $v_{\odot}$ is the system velocity in the heliocentric  
                   system [km~s$^{-1}$]} \\
\multicolumn{6}{l}{$\rm ^c$ $v_{\rm exp}$  the expansion velocity of the AGB 
                    ejecta [km~s$^{-1}$]} \\
\end{tabular}
\newline
B: this study; 
D: Dyck et al 1991;
F: Fernie 1995;
H: Huggins et al. 1988;
K: Kodaira et al. 1970;
J: Jura et al. 1995;
L: Likkel et al. 1987; 
O: Omont et al. 1993;
R: Reddy et al. 1997;
T: Te Lintel Hekkert et al. 1991; 
V: van Winckel et al. 1995;
W: Woodsworth et al. 1990; 
Y: Young et al. 1992; 
Z: Zuckerman et al. 1986
\newline
\end{table*}

In Paper~I we showed that the molecular absorption lines
in the spectrum of HD~56126 are formed in the AGB ejecta. Now we
investigate for the current sample 
if there is a correlation between the expansion 
velocities
of the AGB ejecta derived from CO (or OH for IRAS~08005-2356)
emission line profiles, 
and the velocity differences 
between the molecular absorption lines and the system velocities. The system 
velocity was taken to be the central velocity of the CO or OH line emission 
profile or from radial velocity studies on binaries (Table~4).
The velocities of the photospheric lines are not a good estimate of
the system velocity, since
most of the star in the sample are somewhat pulsating and/or are binaries.

\begin{figure*} % Fig. 4
\centerline{\hbox{\psfig{figure=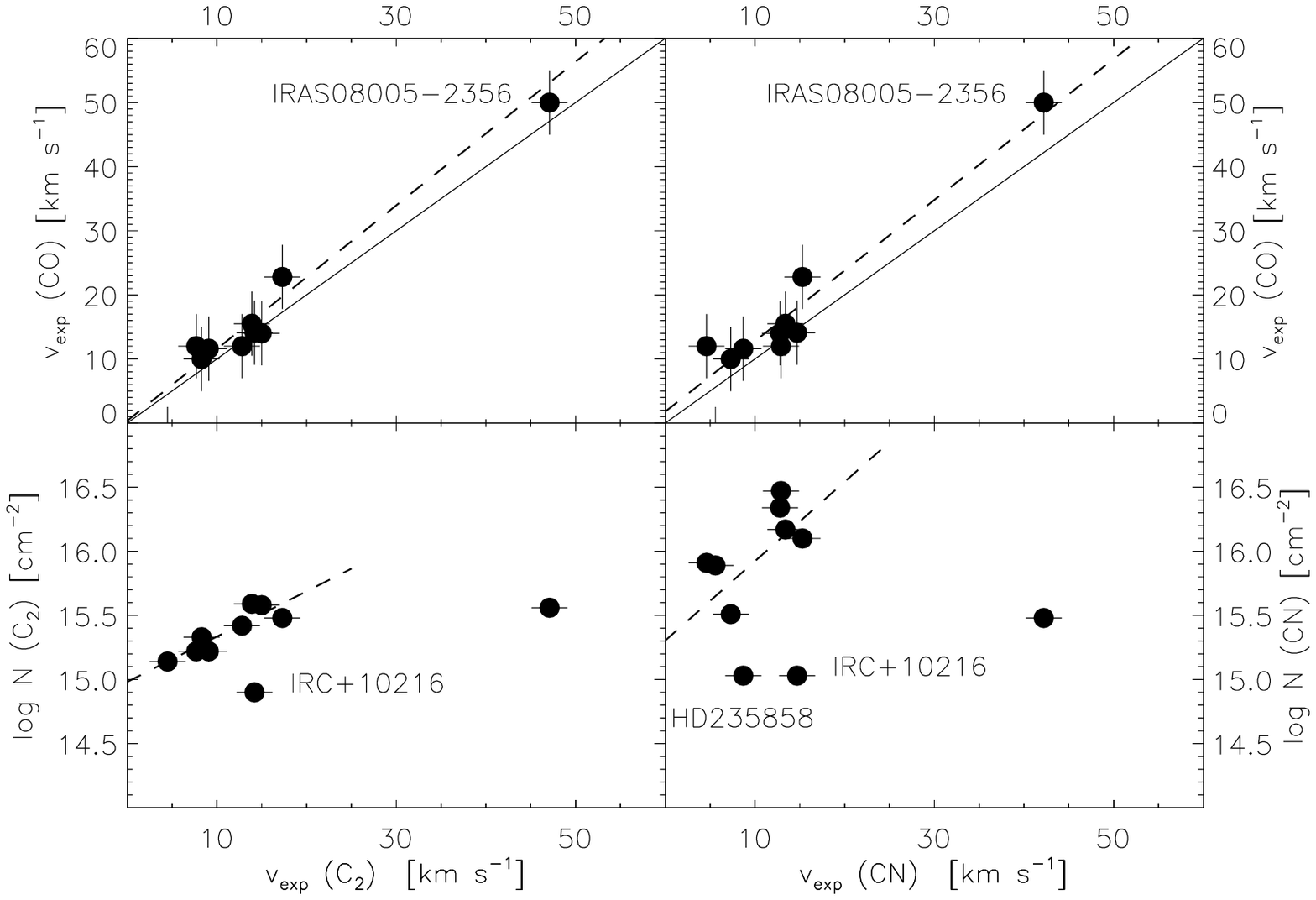,width=\textwidth}}}
\caption{Top panels: expansion velocities derived from CO 
(or OH for IRAS~08005-2356) versus expansion
velocities derived from C$_{2}$ and CN. Lower panels: C$_{2}$ and CN column 
density versus C$_{2}$ and CN expansion velocity.   
The solid lines in the upper two panels give the relations
in which the two velocities are equal. The dashed 
lines are linear least-squares fits as given in the text. IRAS~05341+0852 is not plotted.}
\end{figure*}

\begin{table*} % tab 5
\centerline{\rotate[l]{
\begin{tabular}{lrrrrrrrrrrrcl}
\multicolumn{14}{l}{{\bf Table 5.} Physical parameters derived from the molecular lines.} \\
\multicolumn{14}{l}{~~} \\
\hline
\hline
&\multicolumn{5}{c}{C$_{2}$ $\rm A^{1}\Pi_{u}-X^{1}\Sigma^{+}_{g}$ 
(3,0) or (2,0) $\rm ^a$}
&\multicolumn{5}{c}{CN $\rm A^{2}\Pi-X^{2}\Sigma^{+}$ (2,0) or (3,0) $\rm ^b$}
&  & & \\
Object&$v_{\odot}$&$v_{\rm exp}$                $\rm ^c$
      &$T_{\rm rot}$&$\log N_{\rm mol}$ $\rm ^d$
      &$\log \dot M$
      &$v_{\odot}$&$v_{\rm exp}$                $\rm ^c$
      &$T_{\rm rot}$
      &$\log N_{\rm mol}$               $\rm ^d$
      &$\log \dot M$
      &$\delta v \rm$                   $\rm ^e$
      &$\frac{N \left({\rm CN} \right)}{N \left( {\rm C_{2}} \right)}$
      &$^{12}$C/$^{13}$C \\
      &[km s$^{-1}$] & [km s$^{-1}$]&[K]  &[cm$^{-2}$]&[M$_{\odot}$ yr$^{-1}$]   
      &[km s$^{-1}$] & [km s$^{-1}$]&[K]  &[cm$^{-2}$]&[M$_{\odot}$ yr$^{-1}$]&     &     &       \\
      &             &$\pm2.0$    &      &$\pm0.10$  &$\pm1$           
      &             &$\pm2.0$    &      &$\pm0.10$  &$\pm1$     &      &$\pm2.5$ &       \\
\hline
                 &             &       &     &              &      
                 &             &       &     &              &      &      &      &         \\     
IRAS~04296+3429  &$-66.7\pm0.4$&$  7.7$&$138\pm14$&$15.21(15.20)$&$-5.8$ 
                 &$-63.6\pm0.6$&$  4.6$&$ 25\pm06$&$15.91       $&$-5.2$&$-3.1\pm0.8$&$ 4.9$&$\geq 20$\\
IRAS~05113+1347  &$ -2.5\pm0.6$&$  4.5$&$198\pm31$&$15.13(15.13)$&$-6.1$  
                 &$ -3.6\pm1.1$&$  5.6$&$ 19\pm02$&$15.89       $&$-5.1$&$ 1.1\pm1.3$&$ 5.6$&$\geq 20$\\
IRAS~05341+0852  &$ 13.4\pm0.8$&$ 11.6$&$ 77\pm09$&$15.00(14.61)$&$-6.0$
                 &$ 15.0\pm0.4$&$ 10.0$&$ 38\pm04$&$15.86       $&$-5.0$&$-1.6\pm1.0$&$ 7.2$&\nodata \\
HD~56126         &$ 77.3\pm0.1$&$  8.3$&$242\pm34$&$15.31(15.28)$&$-5.5$
                 &$ 78.3\pm0.2$&$  7.3$&$ 25\pm08$&$15.51       $&$-5.2$&$-1.0\pm0.2$&$ 1.5$&$\geq 20$\\
IRAS~08005-2356  &$ 17.4\pm0.4$&$ 43.7$&$149\pm10$&$15.55(15.60)$&$-4.7$
                 &$ 22.3\pm0.2$&$ 42.2$&$ 38\pm02$&$15.48       $&$-4.7$&$-4.9\pm0.4$&$ 0.8$&$\geq 11$\\
IRC~+10216       &$-33.3\pm0.7$&$ 14.2$&$ 43\pm16$&$14.90(15.15)$&$-6.3$
                 &$-33.8\pm0.5$&$ 14.7$&$ 28\pm05$&$15.03       $&$-6.0$&$-0.5\pm0.9$&$ 1.5$&\nodata  \\
IRAS~20000+3239  &$-16.9\pm0.4$&$ 12.8$&$234\pm33$&$15.41(15.44)$&$-5.3$
                 &$-17.0\pm0.3$&$ 12.9$&$ 50\pm02$&$16.47       $&$-4.1$&$ 0.1\pm0.5$&$11.2$&$\geq 20$\\
AFGL~2688        &$-66.5\pm0.3$&$ 17.3$&$ 56\pm08$&$15.47(15.51)$&$-4.8$  
                 &$-64.5\pm0.5$&$ 15.3$&$ 18\pm03$&$16.10       $&$-4.1$&$-2.0\pm0.6$&$ 4.2$&$\geq 19$\\
IRAS~22223+4327  &$-57.2\pm0.2$&$ 15.0$&$399\pm36$&$15.57(15.55)$&$-4.6$ 
                 &$-55.0\pm0.3$&$ 12.8$&$ 30\pm04$&$16.34       $&$-3.8$&$-2.2\pm0.4$&$ 5.8$&$\geq 20$\\
HD~235858        &$-52.2\pm0.3$&$  9.1$&$119\pm35$&$15.20(15.27)$&$-5.7$ 
                 &$-51.8\pm0.1$&$  8.7$&$ 24\pm17$&$15.03       $&$-5.8$&$-0.4\pm0.3$&$ 0.6$&$\geq 11$\\
IRAS~23304+6147  &$-39.7\pm0.3$&$ 13.9$&$281\pm21$&$15.57(15.57)$&$-5.0$
                 &$-39.2\pm0.5$&$ 13.4$&$ 43\pm14$&$16.17       $&$-4.3$&$-0.5\pm0.6$&$ 3.8$&$\geq 20$\\
                 &             &       &     &              &      
                 &             &       &     &              &      &      &      &         \\     
\hline
&\multicolumn{5}{c}{CH$^{+}$ $\rm A^{1}\Pi-X^{1} \Sigma^{+}$(0,0)} \\
\hline
                 &             &       &     &              &      
                 &             &       &     &              &      &      &      &         \\     
HD~44179$\rm ^f$ &$ 17.3\pm0.5$&$ 3.4$&$202\pm16$&$14.44(14.49)$&\nodata 
                 &\nodata      &\nodata&\nodata&\nodata     &\nodata&\nodata&\nodata&$\geq 22$\\
HD~213985        &$-52.4\pm0.6$&$ 6.7$&$155\pm22$&$14.35(14.35)$&\nodata 
                 &\nodata      &\nodata&\nodata&\nodata     &\nodata&\nodata&\nodata&$\geq 38$\\
                 &             &       &     &              &      
                 &             &       &     &              &      &      &      &         \\     
\hline
\hline
\multicolumn{14}{l}{$\rm ^a$ C$_{2}$ (2,0) for IRC~+10216, else C$_{2}$ (3,0)}\\
\multicolumn{14}{l}{$\rm ^b$ CN (2,0) for IRAS~08005-2356, HD~235858, and IRC~+10216, 
                             else CN (3,0)} \\
\multicolumn{14}{l}{$\rm ^c$ $v_{\rm exp}=v_{\rm sys}  -  v_{\rm C_{2}/CN/CH^{+}} $} \\
\multicolumn{14}{l}{$\rm ^d$ in brackets: sum over $J''$ or $N''$ levels of observed  
                             transitions} \\
\multicolumn{14}{l}{$\rm ^e$ $\delta v=  v_{\rm CN}  - v_{\rm C_{2}}$} \\
\multicolumn{14}{l}{$\rm ^f$ the emission line spectrum has been analyzed with A$^{1}\Pi$ $J'=0$
                             as energy zerolevel.} \\
\multicolumn{14}{l}{$v_{\rm exp}$ and $T_{\rm rot}$ are real, but
                             the column density of HD~44179 is rather meaningless} \\
\end{tabular}
}}
\end{table*}

The upper panel of Fig.~4 shows that the expansion velocity derived from 
CO or OH emission lines and those derived from C$_{2}$ or CN absorption lines
are identical within the estimated error of 2~km s$^{-1}$. The expansion velocities
determined from C$_{2}$, CN, CH$^{+}$ are tabulated in Table~5.
This proves unambiguously that the 
line-forming region of these molecular absorption lines is the AGB ejecta
(circumstellar shell). 
The optical molecular lines are not resolved ($FWHM \sim 6.0$~km s$^{-1}$), which puts
an upper limit of 6.0~km s$^{-1}$ on the turbulent broadening and velocity 
stratification within the line-forming region.
The CH$^{+}$ detections are discussed in \S 4,
but to first order they have the same range of expansion velocities.
A linear least-squares fit (excluding IRAS~05113+1347, IRAS~05341+0852 for
which no CO data are available) gives:

\begin{eqnarray}
\label{eq-coc2}
v_{\rm exp} \left({\rm CO}   \right) = 1.12 \times 
v_{\rm exp} \left({\rm C_{2}}\right) + 0.36 \\
\label{eq-cocn}
v_{\rm exp} \left({\rm CO}   \right) = 1.10 \times 
v_{\rm exp} \left({\rm CN}   \right) + 1.80 
\end{eqnarray}

\noindent
with correlation coefficients of 0.98 and 0.97, respectively.
All velocities are in units of km s$^{-1}$.
\noindent

\subsubsection{Rotational temperature and molecular column density}

The second step is to determine the equivalent widths of the lines and to 
construct rotational diagrams (Fig.~5, 6, 7, and 8). All rest wavelengths,
(absorption) oscillator strengths and equivalent widths of the lines used are tabulated in
App.~A (available only at CDS or from the authors).
The adopted method is extensively
discussed in Paper~I, and we refer the interested reader
to that paper for all details on the 
rotational diagram. Unfortunately there is a error in Eq.~5 of Paper~I.
The correct equation is: $S^{\rm Q}_{J''} = (2 J''+1)$ (page 208, Herzberg 1950).

\begin{figure*} % Fig. 5
\centerline{\hbox{\psfig{figure=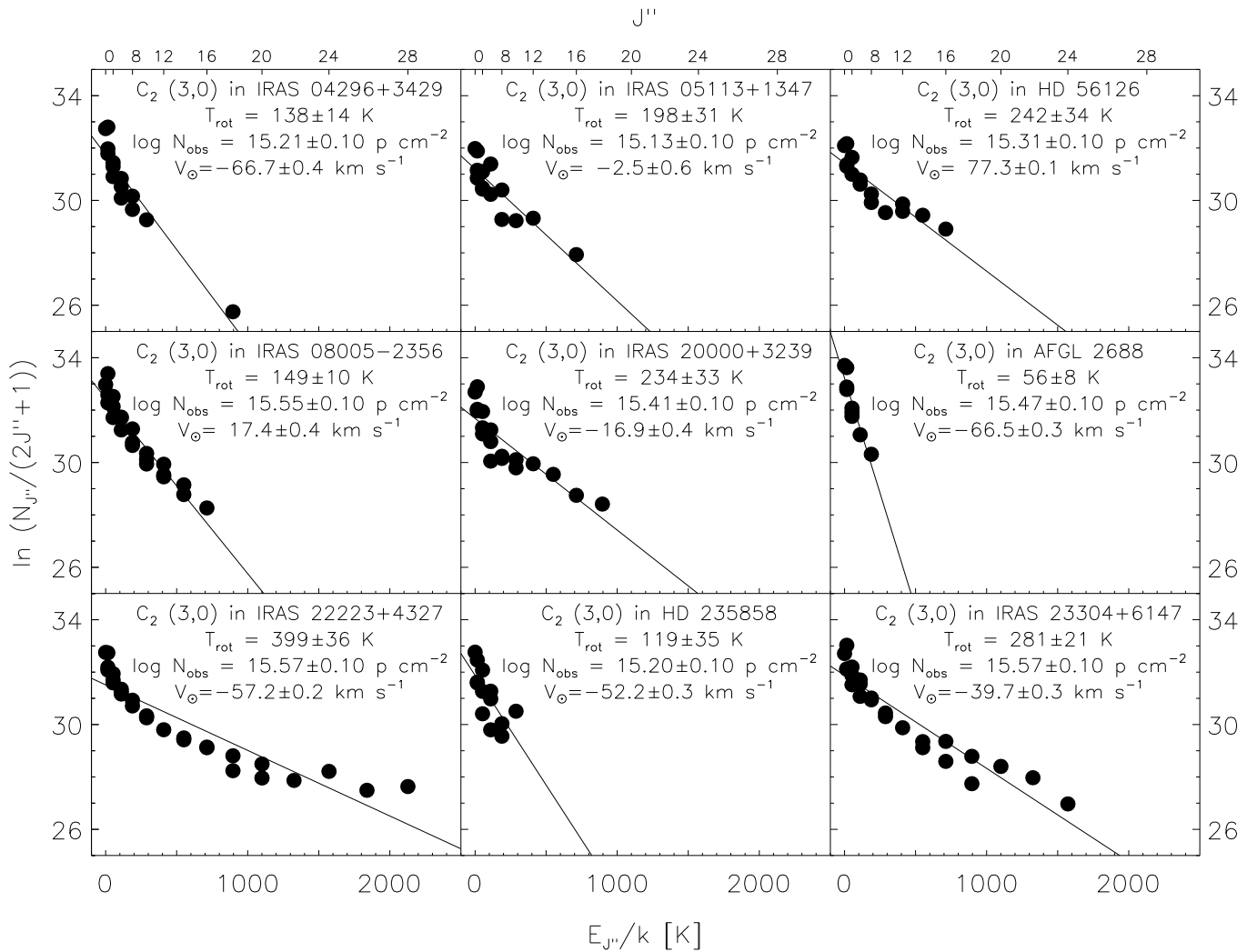,width=\textwidth,rheight=14cm}}}
\caption{Rotational diagrams for the stars showing the Phillips C$_{2}$ 
(3,0)  absorption. Note that for those stars for 
which transitions from high $J''$ levels are  observed (e.g., IRAS~22223+4327), 
the rotational diagram shows a flattening  due to optical pumping of the 
molecule (see \S 3.2.2 for details).}
\end{figure*}

\begin{figure*} % Fig. 6
\centerline{\hbox{\psfig{figure=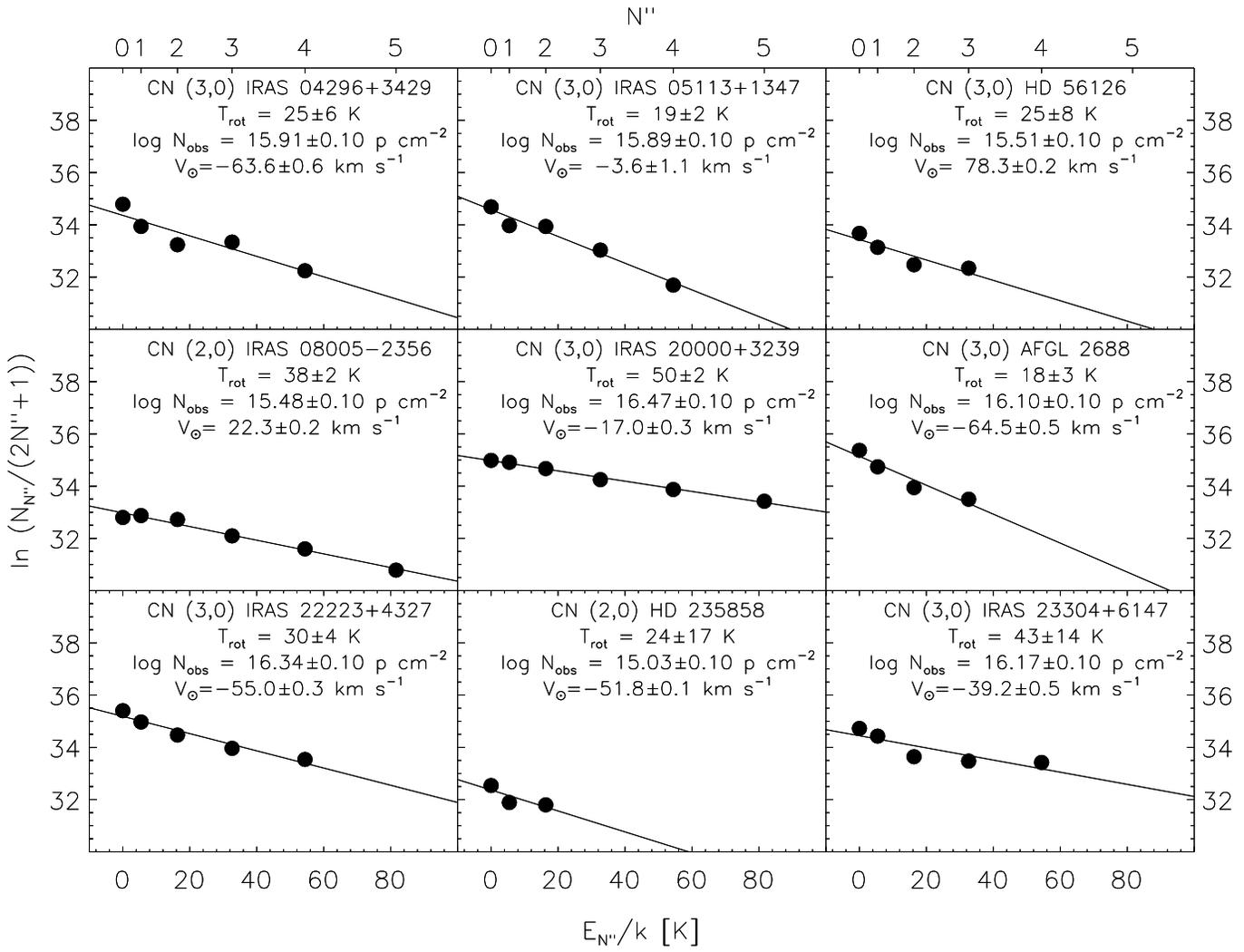,width=\textwidth,rheight=14cm}}}
\caption{Rotational diagrams of the CN Red System (3,0) band 
(if not detected the (2,0) band) for the
stars showing CN absorption. Note that the energies have been computed
relative to $N''=0$.}
\end{figure*}

\begin{figure*} % Fig. 7
\centerline{\hbox{\psfig{figure=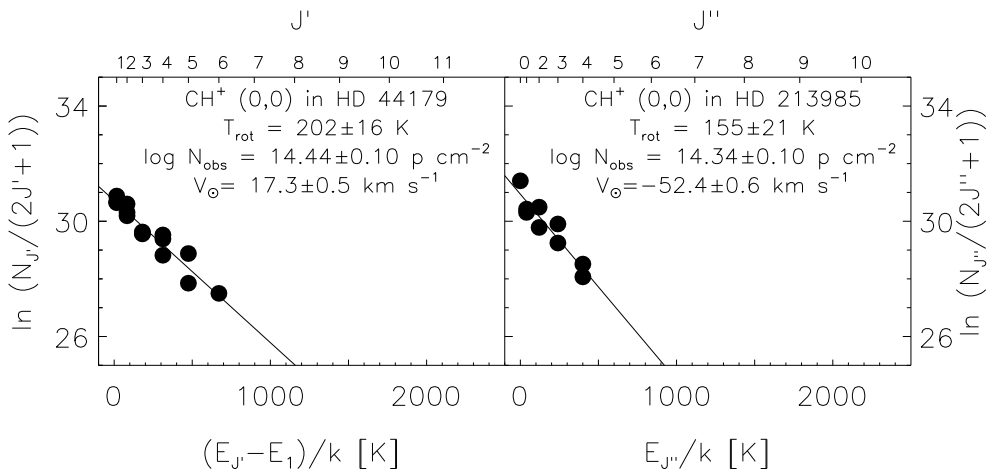,width=\textwidth,rheight=6cm}}}
\caption{Rotational diagrams for the stars showing the CH$^{+}$
(0,0) absorption or emission. For CH$^{+}$ emission the energy levels
are relative to A$^{1}\Pi$ $J'=1$.}
\end{figure*}

\begin{figure*} % Fig. 8
\centerline{\hbox{\psfig{figure=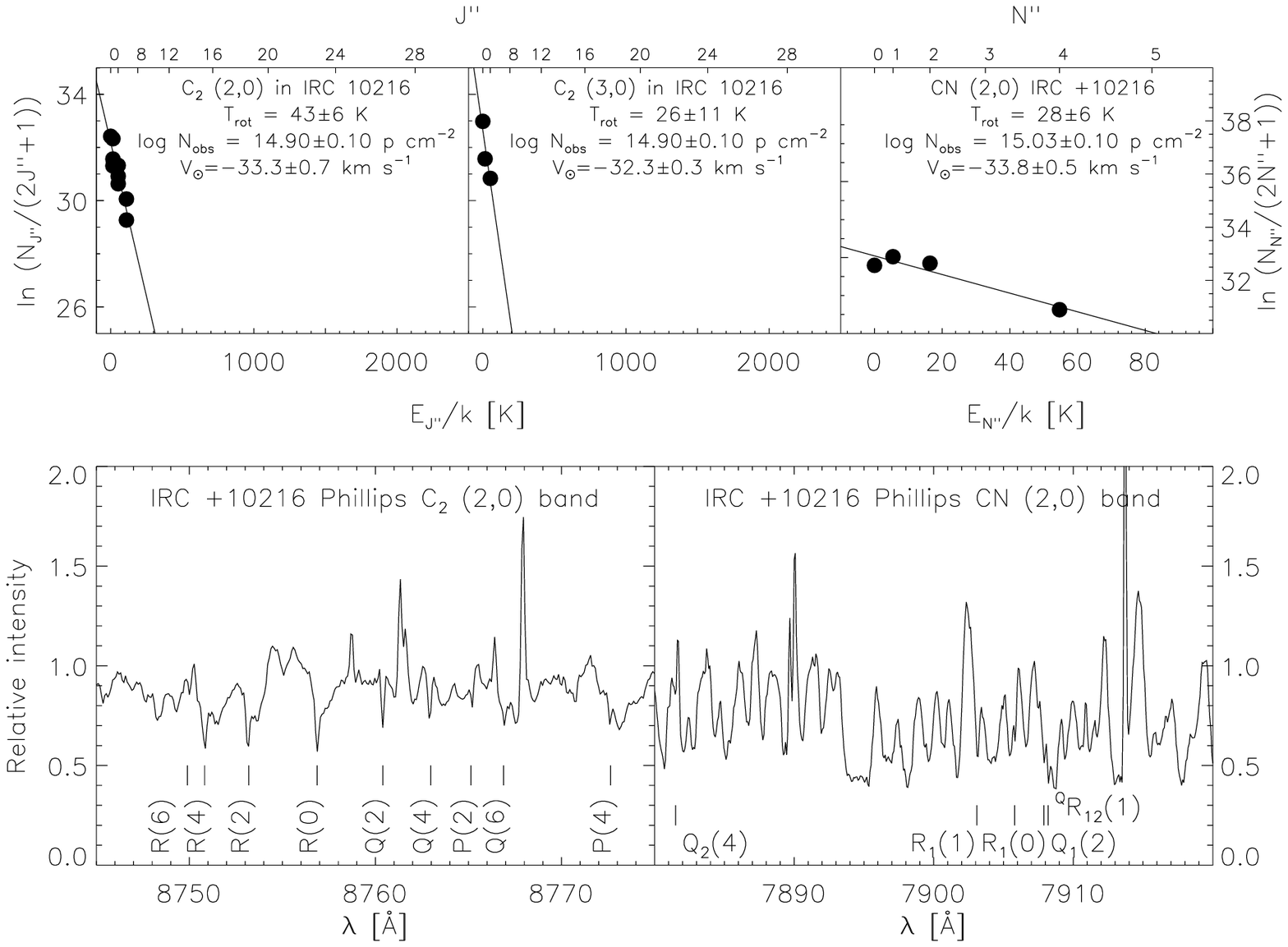,width=\textwidth}}}
\caption{The rotational diagrams of the  C$_{2}$ Phillips (2,0) 
and (3,0) and CN Red System (2,0) absorption bands
in the spectrum of IRC~+10216. The lower  panels gives the part of the 
spectrum on which we have based our firm identification of C$_{2}$ and 
tentative identification of CN (see text for details).}
\end{figure*}

The (absorption) oscillator strengths were computed in the usual way (see Paper~I 
for further references) using  band oscillator strengths of  
$f_{(2,0)}=1.44\times10^{-3}$ and
$f_{(3,0)}=6.672\times10^{-4}$ for C$_{2}$ Phillips (2,0) and
(3,0) band respectively (Langhoff et al. 1990, Langhoff 1996),
and $f_{(0,0)}=5.45\times10^{-3}$ 
for CH$^{+}$ (0,0) (Larsson \& Siegbahn 1983).
For CH$^{+}$ emission the emission oscillator strength is given
by $g_{J'} f_{emission} = g_{J''} f_{absorption}$.
All data of the CN Red System were extracted from the SCAN tape of 
J{\o}rgenson \& Larsson (1990) with the oscillator
strength multiplied by 0.734 as suggested by the authors.

The rotational temperature is determined using rotational constant,
$B_{v}$, values of 1.8111, 1.8907, 13.9302, and
11.4227 cm$^{-1}$ for  
C$_{2}$ ${\rm X^{1}\Sigma^{+}_{g}}~v'' =0$ (Marenin \& Johnson 1970), 
CN ${\rm X^{2}\Sigma^{+}}~v''=0$ (Brocklehurst et al. 1971),
CH$^{+}$ ${\rm X^{1}\Sigma^{+}}~v''=0$  (Carrington \& Ramsay 1982)
and  CH$^{+}$ ${\rm A^{1}\Pi}~v'=0$ (Carrington \& Ramsay 1982), respectively.
Higher order corrections terms (e.g. D$_{v}$, H$_{v}$) where included if needed.

In Paper~I we have shown that only the C$_{2}$ and
CN (3,0) band of HD~56126 are optically thin. In order to
facilitate the analysis, we have limited ourself to the C$_{2}$ (3,0) and CN (3,0) bands.
For each molecular band we investigated whether the different branches
for a given lower level gave the same column density. Within the errors
of the determination of the equivalent width this was indeed the case.
In case the lines would have been optically thick, this would
have shown up as vertical scatter in the 
rotational diagrams (Fig.~5,  and 6). 
This is not observed, and it confirms that the lines are optically
thin.

Recently Bakker et al. (1996d) studied
the C$_{2}$ and CN bands of HD~56126 by means of a curve of growth 
analysis. They find that the (3,0) bands are indeed close to being
optically thin. However, in this work any errors which are introduced due
to the assumption that the lines are optically thin are negligible with respect to
the errors in the measurements of the equivalent width.

CH$^{+}$ absorption and emission is assumed to be optically thin
based on the narrow correlation of the data points in the rotation diagram
(Fig.~7), and the fact that different branches (P, Q, and R) do yield
comparable column densities for the $J''$ (or $J'$) level they originate from.
The relative rotational diagram for CH$^{+}$ emission is relative to
the energy level of the A$^{1}\Pi$ $J'=1$ level. This energy level
is at 42400~K (over forty thousand degrees !) above ground level.

If the population distribution over the rotational levels
follows a Boltzmann distribution, then the rotational diagram should be a linear
relation for optically thin lines  ($\tau \leq 1$). 
The slope of the curve is inversely proportional to the rotational  temperature
$T_{\rm rot}$ and the offset is given by the natural logarithm of  
$N_{\rm mol}/Q_{r}$, where $N_{\rm mol}$ is the column density of the molecule 
and $Q_{r}$ the partition function. 
If the rotational temperature is not equal to the kinetic temperature and different
for each pair of levels, then 
the population distribution over
the rotational levels is non-Boltzmann and the rotational diagram 
will be non-linear.
From linear least-squares fits to the rotational diagrams, we find 
typical values  of 
$T_{\rm rot}=43-399$~K, $\log N_{\rm mol}=14.90-15.57$~cm$^{-2}$ 
for C$_{2}$ (3,0) (or (2,0)),
$T_{\rm rot}=155-202$~K, $\log N_{\rm mol}=14.35$~cm$^{-2}$  
for  CH$^{+}$ (0,0), and 
$T_{\rm rot}=18$ to $50$~K, $\log N_{\rm mol}=15.03$ to $16.47$~cm$^{-2}$ 
for CN (3,0) (or (2,0)) with an average particle ratio of 
$N({\rm CN})/N({\rm C_{2}})= 4.0\pm3.0$

Since the rotational temperature is determined using all available data 
points, it is an average temperature. For C$_{2}$, use of only the 
low $J''$ levels (which are most sensitive to collisional (de-)excitation)
results in a lower rotational temperature, which will be  closer to the 
kinetic temperature of the gas. On the other hand, using only the high-$J''$ 
levels (which are sensitive to radiative pumping)
results in a rotational temperature which will be closer to the color temperature of 
the exciting stellar radiation field.

The column densities are computed using the partition function, which 
depends on the average rotational temperature. In Paper~I we investigated 
the  accuracy of this method and found that the column densities are
slightly underestimated for super-thermal C$_{2}$, and slightly 
overestimated for sub-thermal CN. This introduces a relative 
error of about 10\%.  A slightly more accurate determination of the molecular 
column density could be obtained by adding the column densities of 
all observed $J''$ levels and extrapolating to the unobserved levels.
The oscillator strengths are not very well known and
we adopt an absolute error of  15\%. As a consequence, we will adopt an absolute
error of 15\% and a relative error of 10\%  to the molecular column densities. 

The detection of C$_{2}$ and CN in the same spectrum of the same star with 
significantly different rotational temperatures can be attributed to the
difference in the dipole moment of the two molecules. C$_{2}$ is a homonuclear 
molecule (the electronic configuration is spatially symmetric
with respect to the rotational axis, and the
nuclear spin of $^{12}$C is $I=0$). 
Thus, without a permanent dipole moment the 
selection criteria forbid pure rotational and rotational-vibrational 
transitions in the ground electronic state (except through weak quadrupole transitions). 
A homonuclear molecule has therefore no efficient transitions available to 
emit a photon. 
The higher-$J''$ levels will be overpopulated relative to a Boltzmann distribution at the 
local kinetic temperature, because they can be populated by radiation with a color temperature
larger than the local kinetic temperature (optical pumping). The molecule is
super-thermally excited. In
Fig.~5 this shows up as a flattening of the curve for $J'' \geq 20$ levels.
CN is a heteronuclear molecule and selection criteria allow dipole 
transitions: energy can be easily released and the molecule 
cools to sub-thermal values if the density is below the density for
collisional excitation.

Remarkably, there seems to be a relation between the molecular column density
and the expansion velocity. Since CN can only be photodissociated by
far UV photons ($\lambda \leq 1100$ \AA), it is more sensitive to the interstellar
UV radiation field, this
leads to a large spread in the observed CN column densities.
A linear least-squares fit (excluding IRAS~08005-2356, IRAS~05341+0852, and IRC~10216) yields:

\begin{eqnarray}
\label{eq-nc2}
\log N \left({\rm C_{2}}\right) = 3.53 \times 10^{-2} \times 
v_{\rm exp} \left({\rm C_{2}}\right)  + 14.98  \\
\label{eq-ncn}
\log N \left({\rm CN   }\right)  = 6.19 \times 10^{-2} \times 
v_{\rm exp} \left({\rm CN   }\right)  + 15.3 
\end{eqnarray}

with correlation coefficients of 0.89 and 0.53, respectively.
Column densities and velocities are in units of cm$^{-2}$ and km s$^{-1}$ 
respectively.
A detailed discussion
of this relation will be given in \S~4, where we will argue that it
reflects differences in intrinsic carbon abundance.

\subsubsection{Time scales of AGB ejecta}

The far-infrared  radiation (IRAS data) is due to thermal radiation 
of dust in the AGB ejecta. 
From the IRAS 12 to 25~$\mu$m flux ratio, $f_{v} \left( 12 \mu 
{\rm m} \right) / f_{v}\left( 25 \mu {\rm m} \right)$, we derived the color 
temperature of the innermost dust using the method described in the  IRAS 
Explanatory Supplement (1986) and Table Suppl. VI.C.6. (Table~1).
For stars with a near-infrared excess (e.g., a circumbinary disk) 
this method cannot be applied, and
we cannot calculate the mass-loss rate as described here.

Trams (1991) has fitted an  optically thin dust model to a dozen 
post-AGB stars and found that for all stars the dust emissivity efficiency,
$Q \left( \nu \right) = Q_{0} \left( \nu / \nu_{0} \right) ^{p} $,
has an index parameter $p=1$. Following Sopka et al. (1985), the dust inner radius 
($r_{o}$) is given by:

\begin{equation}
\label{art9eq-rinner}
\frac{r_{0}}{R_{\ast}} = \left( \frac{T_{\rm dust}}{0.6 T_{\rm eff}} \right) ^
{- \left( \frac{4 + p}{2} \right)}.
\end{equation}

\noindent
with $T_{\rm dust}$ and $T_{\rm eff}$ the Blackbody dust temperature and the stellar
effective temperature, respectively.
Adopting  $\log L_{\ast} = 3.86$ $\log$ L$_{\odot}$  for a 0.6~M$_{\odot}$ post-AGB star
(Bl\"{o}cker 1995) the stellar radius, $R_{\ast}$, can be calculate.
Typical distances derived are of the order of $10^{16}$~cm (Table~6). 
Taking the dust inner radius and the expansion velocity, we can subsequently
estimate the time since the star left the AGB (Eq.~\ref{art9eq-timepagb}, Table~6)
and the average  annual increase in effective temperature (Eq.~\ref{art9eq-speed}, Table~6):

\begin{eqnarray}
\label{art9eq-timepagb}
t_{\rm post-AGB} = \frac{r_{0}}{ v_{\rm exp} } \\
\label{art9eq-speed}
\frac{\Delta T}{\Delta t} = \frac{T_{\rm eff} - T_{\rm AGB}}{t_{\rm post-AGB}}
\end{eqnarray}

\noindent
while taking $T_{\rm AGB}=3500$~K.

The stars in our sample left the AGB typically 300 years ago and they have a typical
annual effective temperature increase of 5~K. Since an abundance analysis allows the
determination of the effective temperature as accurate as 100 K (for supergiants),
this suggests that we expect to see change in the effective temperature and
conditions of the molecules on a time scale of about 20 years.

\begin{table*} % tab 6
\caption{Derived stellar and dust parameters assuming $\log L_{\ast}=3.86$ 
$\log$ L$_{\odot}$.}
\begin{tabular}{lrrrrrrrl}
\hline
\hline
Object&$T_{\rm eff}$&$R_{\ast}$& $T_{\rm dust}$& $r_{0}$& $\log r_{0}$&
$t$& $\Delta T/\Delta t$& Remark \\
   & ~[K]& [R$_{\odot}$]& [K]& [$R_{\ast}$]& [cm]& [yr]& [K yr$^{-1}$]& \\
\hline
               &         &       &   &       &       &       &       &                  \\
IRAS~04296+3429& 5550    & 92    &195&1200   &15.89  &320    & 6     &\nodata           \\
IRAS~05113+1347& 4600    &134    &185& 860   &15.91  &570    & 2     &\nodata           \\
IRAS~05341+0852& 6500    & 67    &235&1120   &15.72  &154    &19     &\nodata           \\
HD~44179       &10300    & 27    &320&\nodata&\nodata&\nodata&\nodata&near-ir excess    \\
HD~56126       & 6900    & 60    &175&2720   &16.06  &440    & 8     &\nodata           \\
IRAS~08005-2356& 6900    & 60    &210&1730   &15.86  & 50    &70     &near-ir excess ?? \\
IRC~10216      & 2200    &\nodata&240&\nodata&\nodata&\nodata&\nodata&\nodata           \\
IRAS~20000+3239& 4600    &134    &175& 990   &15.97  &330    & 5     &\nodata           \\
AFGL~2688      & 6900    & 60    &145&4360   &16.26  &330    &10     &\nodata           \\
IRAS~22223+4327& 5550    & 92    &120&4060   &16.42  &550    & 4     &\nodata           \\
HD~235858      & 4850    &120    &185& 980   &15.92  &290    & 5     &\nodata           \\
HD~213985      &10300    & 27    &380&\nodata&\nodata&\nodata&\nodata&near-ir excess    \\
IRAS~23304+6147& 5200    &105    &170&1440   &16.03  &240    & 7     &\nodata           \\
               &         &       &   &       &       &       &       &                  \\
\hline
\hline
\end{tabular}
\end{table*}

\subsubsection{Mass-loss rate}

Here we will attempt to derive the mass-loss rates from the column densities
and expansion velocities in the AGB ejecta. In order to do so, we will make two
important simplifications:

(i) Since there is no simple, unique way to determine the distance 
from the star to the molecules, we will assume that the molecules
reside where dust is present. 
In the case of the extended envelope of AGB stars and in the circumstellar 
shells of post-AGB stars (F to G-type stars),
simple molecules are formed out of complex molecules by photodissociation  
by the interstellar radiation field.
However, for circumstellar shells the  abundance of a simple molecule will 
peak at a distance larger than the dust inner radius. Therefore, 
the assumption that the molecules are formed at the dust 
inner radius and beyond underestimates the derived mass-loss rate.

(ii) Our second assumption is that the molecular abundances are
the same for each star, resemble those of the AGB star IRC~+10216,
and do not change with distance to the star. From our finding of a relation 
between the observed molecular column density and the expansion velocity we 
know that this is not the case. Furthermore, we take the standard abundances 
to be the computed peak abundance for IRC~+10216.
No accurate prediction for molecular abundance of CH$^{+}$ in 
AGB ejecta was found in the literature, while, in the interstellar medium
the large observed abundance of CH$^{+}$ is still not well understood
(see Gredel et al.  1993 for a recent summary).

Using assumptions (i) and (ii) we will derive an equation for
the mass-loss rate in a rather simple manner.
Starting with the general  formula for conservation of mass, 
$\dot M \left( r \right) = 4 \pi r^{2} \rho (r)  v_{\rm exp}  (r) $,
and assuming a constant mass-loss rate over time, $\rho(r)=\rho_{0} \left( r_{0}/r \right)^{2}$,
we find for the density at the dust inner radius::

\begin{equation}
\label{art9eq-rho}
\rho_{0} = \frac{N_{\rm mol}}{r_{0}} . \frac{\mu m_{p}}{X_{\rm mol}}
\end{equation}

\noindent
where $\mu$ is the average hydrogen  (molecular, atomic, and ionic) particle 
mass in units of $m_{p}$ and $N_{\rm mol}$ the molecular column density.  
For the densities and temperatures expected in the 
AGB ejecta most hydrogen will be in the form of H$_{2}$ ($\mu=2.0$). 
In the model of Cherchneff et al. (1993) the molecular 
particle abundances relative to $n_{H}=n({\rm H})+2 \times
n({\rm H}_{2}$) are $X_{\rm C_{2} }=4\times10^{-6}$ and  
$X_{\rm CN}=3 \times10^{-6}$.
Combining Eqs.~5 \& 8 yields the expression for the mass-loss rate:

\begin{eqnarray}
\dot M = \left( 4 \pi \mu m_{p} R_{\odot} 
T_{\odot}^{2} \sqrt{\frac{L_{\ast}}{L_{\odot}}} \right)
. \frac{N_{\rm mol}}{X_{\rm mol}}
. \left( \frac{T_{\rm dust}}{0.6} \right)^{- \left( \frac{4+p}{2} \right)} .
  \nonumber \\
 \left( T_{\rm eff} \right) ^{\frac{p}{2}}  . v_{\rm exp}  =
  \nonumber \\ 
3.71 \times10^{-29} .\frac{N_{\rm mol}}{X_{\rm mol}}. 
T_{\rm dust}^{-\frac{5}{2}} . v_{\rm exp} . \sqrt{T_{\rm eff}}
\end{eqnarray}

\noindent
with all parameters in cgs units except $\dot M$ which is in
M$_{\odot}$ yr$^{-1}$:

\noindent
The mass-loss rates (assuming $\log L_{\ast} = 3.86$ $\log$ L$_{\odot}$)
derived in this manner
(with the theoretical abundances of IRC~10216) are of the 
order  of $-6.2\leq \log \dot M \leq-4.1$ (Table ~5), which
is of the same order of magnitude as derived from the CO emission
and the IR excess.
In view of the important assumptions made in the calculation of the mass-loss
rate we adopt an estimated error of one order of magnitude.
We want to stress again that the assumption made in calculation the mass-loss
rate from the molecular absorption lines only allow an order of magnitude
estimate of the mass-loss rate. We want above all to show that these
molecular lines do allow the determination of the mass-loss rate given
the right inner and outer radius and the molecular abundance. If the 
mass-loss rate is know (from e.g., CO or infrared) 
the process can be reversed and
this would yield the molecular abundance.

A linear least-squares fit (excluding IRAS~08005-2356, IRAS~05341+0852, and IRC~+10216 for
C$_{2}$ and CN and HD~235858 for CN) yields:

\begin{eqnarray}
\label{eq-mdotc2}
 \log \dot M \left( {\rm C_{2}} \right) 
 = 0.11 \times v_{\rm exp} \left({\rm C_{2}}\right)   -6.71  \\
\label{eq-mdotcn}
 \log \dot M \left( {\rm CN   } \right)  
=  0.13 \times  v_{\rm exp}  \left({\rm CN   }\right)  -5.86
\end{eqnarray}

\noindent
with correlation coefficients of 0.95 and 0.92, respectively.
The mass-loss rate and velocity are in units of [M$_{\odot}$ yr$^{-1}$] and
km s$^{-1}$ respectively.

\section{Discussion}

\subsection{General discussion}

We discuss the following points:

1. Within our sample of stars,
eleven stars exhibit C$_{2}$ and CN absorption, one CH$^{+}$ absorption, and one CH$^{+}$ 
emission. All post-AGB stars with the Phillips bands in absorption also show the Swan 
bands in absorption. Since the ${\rm a^{3}\Pi_{u}}$ state is 612~cm$^{-1}$ 
above the ground state, corresponding to $J''=18$ in the ${\rm X^1\Sigma_g^+}$
state, it is likely that this state is populated by optical pumping.
A study of the Swan band will be subject of a separate paper.

2. We note that all the detected molecular bands are clearly detected, while the
non-detections have no molecular bands or their strength is below our 
detection limit. 
It therefore seems that there is no star with intermediate strength molecular 
bands.
Whether this is due to small number statistics or gives clues about 
circumstellar
chemistry is to be evaluated in follow-up research. A suggestion 
for the latter
option could be that the CSE is rather clumpy and that only clumps having a 
density higher 
than a critical  value are able  to sustain the presence of these molecules. 
On the
other hand one might suggest that these molecular bands are rather common, 
but that our 
sample is too small to include stars with molecular bands of intermediate 
strength.

3. All stars exhibiting the unidentified 21~$\mu$m infrared features show 
the presence 
of C$_{2}$ and CN absorption, but this is not true for the reverse. The 
exception, 
IRAS~08005-2356, has the highest column density of C$_{2}$ in  our sample, 
and exhibits OH maser emission at the same velocity.
At the time of the observation, IRAS~05341+0852, was the only 21~$\mu$m 
sources not included in our sample. A recent study on the optical
spectrum of this star by Reddy et al. (1997) has indeed show the
presence of circumstellar C$_{2}$ and CN absorption.

We therefore predict that the recently found 
Post-AGB 21~$\mu$m sources 
(IRAS~22574+6609, Hrivnak \& Kwok 1991a, 
 IRAS~15553-5230 and IRAS~17195-2710, Henning et al. 1996, 
 SAO~163075, Justtanont et al. 1996) 
will show C$_{2}$ and CN absorption.
Whether the other new candidate and waiting list 21~$\mu$m stars
 (YSO and HII regions)  listed
by Henning et al. show molecular absorption, needs to be investigated.
But if they do, then there is a one-to-one relation between the occurrence
of the 21~$\mu$m emission feature and the presence of C$_{2}$ and CN absorption.

We did not found any clear correlation between the parameters listed
in Table~5 and the strength of the 21~$\mu$m feature (Justtanont et al. 1996,
Henning et al. 1996). Though there is a  suggestion for 
an anti-correlation between the $N({\rm CN})/N({\rm C_{2}})$ and
$I(20 \mu{\rm m})/I(18 \mu{\rm m})$.
Unfortunately, small number statistics do not allow any firm conclusion.

4. The presence of C$_{2}$ and CN or CH$^{+}$  seems to be correlated 
with the dust  color temperature:
$T_{\rm dust}\leq300$~K  and $T_{\rm dust}\geq300$~K respectively.  
C$_{2}$ and CN are correlated 
with cold dust (far-infrared excess) and CH$^{+}$ with hot dust 
(near-infrared excess). This idea is supported by the rotational 
temperature of the  
heteronuclear species: 
$T_{\rm rot}\approx 34$ and 200~K for CN and CH$^{+}$ respectively.

5. The stars in our sample typically left the AGB 300~years ago and 
evolved to the left in the HR diagram with a typical annual temperature 
increase of 5~K. IRAS~08005-2356 has an average annual temperature
increase of 70~K per year. We predict that we can see evolutionary
changes of the star and the CSE in the next 20 years.

6. Traditionally, the expansion velocities of the AGB ejecta are determined from
the full width of the CO($J=1\rightarrow 0$) millimeter line profile. Since CO 
is abundant throughout the whole AGB ejecta, the profile contains the 
integrated 
emission. In the case of expansion velocities derived from line
absorption, e.g., C$_{2}$, CN, and CH$^{+}$, the line-forming region is along
the 
line-of-sight towards the stellar photosphere. A second important difference 
is that the molecules observed in absorption are only present in a thin shell 
of material where the conditions are such that the molecule 
has a large abundance. This 
allows us to study the AGB ejecta at different radii, using different 
molecules.
Unfortunately, from a theoretical study on the extended envelope of
IRC+10216 (Cherchneff et al. 1993), is seems that 
the line-forming regions of C$_{2}$ and CN are almost the  same.
The upper panels of Fig.~4 clearly show the correlation between CO and the 
C$_{2}$  and CN expansion velocity, the relations are 
given in Eqs.~\ref{eq-coc2} and \ref{eq-cocn}.
This unambiguously proves that the molecular  absorption lines originate
in the AGB  ejecta and that these are thus circumstellar.

\begin{figure*} % Fig. 9
\centerline{\hbox{\psfig{figure=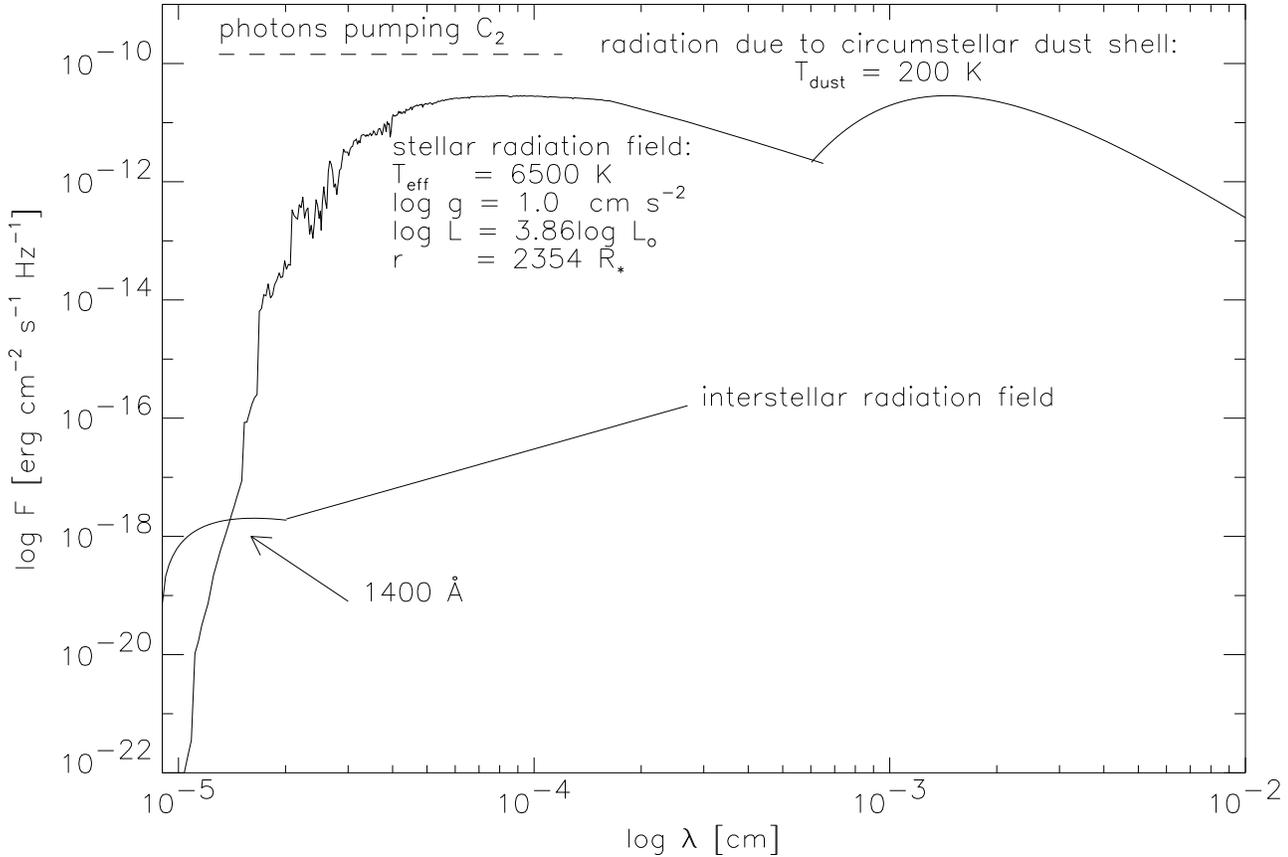,width=\textwidth}}}
\caption{The interstellar radiation field (solid straight line)
and the stellar radiation field at $r_{o}=2719$ R$_{\ast}$  
of a $T_{\rm eff}=6500$~K, $\log g=1$ Kurucz model. 
A typical infrared excess has been visualized by a 
$T_{\rm dust}=200$~K black-body
(scaled arbitrarily). Optical pumping of C$_{2}$ uses photons between
1300~\AA~ and 1.1~$\mu$m (horizontal dashed line), while photodissociation
of C$_{2}$ and CN needs photons with $\lambda \leq 2000$ and $1100$~\AA,
respectively.}
\end{figure*}

7. The lower panels of Fig.~4 show that the C$_{2}$ column density increases 
as a function of expansion velocity (Eqs. \ref{eq-nc2} and \ref{eq-ncn}).
To interpret this result, it is important 
to realize that the line-forming region of C$_{2}$  is only a thin shell.
The inner radius is determined at a critical dust column density 
such that stellar and/or interstellar 
photons can penetrate and photodissociate complex molecules like C$_{2}$H$_{2}$ 
(via C$_{2}$H) and HCN 
into C$_{2}$ and CN, while the outer radius is determined in a similar way
for the photodissociation of C$_{2}$ and CN to individual atoms by the
interstellar and/or stellar radiation field. 
First, we will address the question of which radiation field (stellar,
interstellar, or circumstellar) is responsible for the excitation and
the photodissociation of C$_{2}$ and CN. 
Fig.~9 shows the interstellar radiation field
(Draine (1978) for $\lambda \leq 2000$~\AA~ and van Dishoeck \& Black (1982)
for $\lambda \geq 2000$~\AA), and a $T_{\rm eff}=6500$~K
stellar model (Kurucz 1979) with a $T_{\rm dust}=200$~K dust shell. 
The radiation field of the star is scaled to
a post-AGB star of $\log L=3.86$ $\log$ L$_{\odot}$ observed at a
typical dust inner radius of $r_{o}=2354$ R$_{\ast}$. For wavelengths 
$\lambda \geq 1400$~\AA~ the stellar radiation field dominates, while
for $\lambda \leq 1400$~\AA~ the interstellar radiation field dominates.
Since C$_{2}$ is pumped by optical radiation (1300~\AA~ to 1.1~$\mu$m), 
the stellar radiation 
field  is responsible for the excitation of this molecule. 
Since the CN molecule can only be photodissociated by photons with 
$\lambda \leq 1100$~\AA~, the interstellar ultraviolet radiation field
photodissociates CN. The C$_{2}$ molecule is 
photodissociated by photons with $\lambda \leq 2000$~\AA~ and 
can therefore be dissociated by both the ultraviolet interstellar and
stellar radiation field.

We have clearly found a relation between the expansion velocity
and the column density of C$_{2}$ and to a lesser extend of CN.
The sign of the slope is inconsistent with models which assume
a constant carbon abundance and mass-loss rate: these models
predict a decrease in column density for increasing expansion velocities.
\newline
Alternatively we propose two models: \newline
Our favorite model (model~I): The radiation pressure efficiency
($\overline{Q}_{rp}$ see Tielens 1983) scales with the carbon abundance.
Since the momentum equation gives a positive correlation between 
$\overline{Q}_{rp}$
and the acceleration $dv_{g}/dr$, this would account for the spread in
the observed expansion velocities. At the same time the spread in
carbon abundance in the ejecta will be observable as a spread in the 
observed column densities in a way consistent with our findings.
We ran some simple models and found that only a small increase of 
$\overline{Q}_{rp}$ gave rise to a significant change in expansion velocity.
Additional and much more detailed modeling is needed to determine the exact
relation between $v_{\rm exp}$, $\log N$, and 
$\overline{Q}_{rp}$.
\newline
We propose  an alternative model (model~II):
 The higher the initial mass of the star, the higher the mass-loss
rate and terminal velocity of the wind 
when the star terminates the AGB evolution (Barnbaum et al. 1991).
Since an increase in mass-loss rate increases the column density,
but an increase of the expansion velocity decreases it, the model
can only explain the observation if the increase of mass-loss rate
dominates over the increase of the expansion velocity. Since
the number of thermal pulses increases with the mass of the progenitor,
we also expect an increase of carbon abundance which would work in favor
of this model. Therefore model II included the effect discussed for model I.

Since the model with the least free parameters is stronger than a more
complex model, we favor model~I and  argue that a 
high C$_{2}$ (and CN) abundance reflects a high carbon abundance
(and the C/O ratio),  which in 
turn implies that the stars have experienced different amounts
(number or efficiency) of third dredge-ups. This hypothesis
implies that there should be a relation between the carbon abundance 
of the star
and the observed column density of circumstellar C$_{2}$.

We further note that there seems to be a flattening 
(possibly due to saturation of the absorption lines, flat part
 of curve of growth) in the molecular
column densities for $v_{\rm exp} \geq 30$~km s$^{-1}$,  
but whether this is real or 
an artifact of small number statistics is not clear at this moment.

8. There should be a reasonable explanation why the 
relation between $\log N$ and $v_{\rm exp}$ for CN is not as
well defined as for C$_{2}$.
Although the CN molecular spectra are more complex than the C$_{2}$ spectra,
and a unique identification of an absorption line with a single transition
is not always possible, it is unlikely that this can account for the
large scatter around the mean relation. 
Noting that all stars with $m_{v} \geq 12$ have  
$N$(CN)/$N$(C$_{2}$)$\geq 3$,
while all stars with $m_{v} \leq 12$ have  $N$(CN)/$N$(C$_{2}$)$\leq 3$,
we suggest that the CN abundance might be affected by the distance of 
the star above the Galactic plane, and the local interstellar ultraviolet
radiation field. It might also be that the fainter objects have
a larger circumstellar reddening and therefore the molecules
are more shielded against the stellar/interstellar radiation field.
This would increase $N({\rm CN})$. However, information about
the circumstellar reddening of these objects is  not available.

9. Mass-loss rates have been derived from the observed 
  column densities by making two very important assumptions.
  The basic conclusion is that the mass-loss rates determined are of
  the same order as those obtained from CO emission and infrared measurements,
  but do not allow a detailed comparison between different objects.

\begin{figure*} % Fig. 10
\centerline{\hbox{\psfig{figure=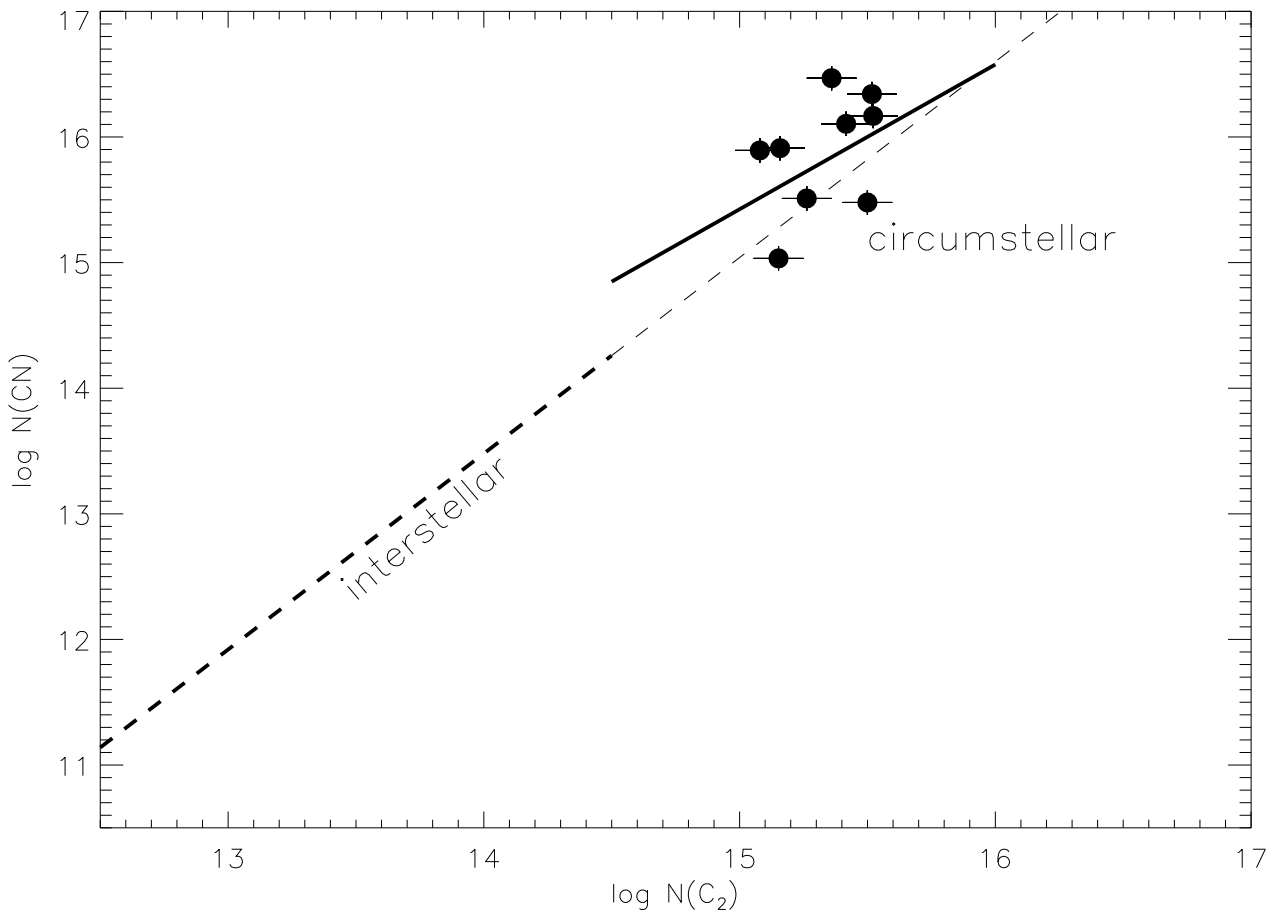,width=\textwidth}}}
\caption{A logarithmic plot of N(CN) versus N(C$_{2}$). 
The thick dashed line represents the interstellar data points from 
Federman et al. (1994), while the circumstellar data points
are determined in this study. Is seems that the
circumstellar points follow the interstellar trend.}
\end{figure*}

10. Federman et al. (1994) presented a relation between the C$_{2}$ 
and CN column
density observed in diffuse interstellar clouds. We have investigated if
this relation could be extrapolated to the conditions 
prevailing in the CSE of post-AGB stars (Fig.~10). 
We find (excluding IRC~+10216, and IRAS~05341+0852):

\begin{eqnarray}
\log N \left( {\rm CN} \right) = 1.153 \times \log N \left( {\rm C_{2}} \right) -1.863
\end{eqnarray}

\noindent
with $N$ in units of cm$^{-2}$ with a low correlation coefficient of 0.43.
Although both in the interstellar and circumstellar case C$_{2}$ and CN
are primarily destroyed by photodissociation by the interstellar radiation
field, the formation chemistry of these species is assumed to be very
different in interstellar clouds.

11. We have searched for the lower $v'$ bands (since they are the strongest);
we did not detect a single isotopic line of C$_{2}$, CN, or CH$^{+}$.  
Depending on the spectrum
the detection limit is about 5 to 10 m\AA, which gives a typical lower limit 
of 
$\rm ^{12}C/^{13}C >20$. $^{13}$C is produced by the CN-cycle and not by
the triple-$\alpha$ process. A high isotope ratio indicates that the carbon
enhancement is due to the third-dredge up (convection reaches the deeper 
He-burning shell). 
This is rather surprising since one would naively expect that the material 
from the outer shell
(H-burning) can be dredged-up easily. The high isotope ratios of these stars 
is consistent with carbon stars (AGB stars) to be the progenitor 
of the stars studied 
($^{12}$C/$^{13}$C $\approx 40 - 80$, Lambert et al. 1986).
The third dredge-up only occurs on the thermal pulsating
AGB phase for stars with $M_{\rm MS} \leq 5$~M$_{\odot}$
($Z=0.001$): this sets an upper limit
on the star's initial main sequence mass. For higher initial masses, 
Hot Bottom Burning 
(HBB) will prevent the formation of a carbon rich post-AGB star 
(Boothroyd et al. 1993).

12. Ultraviolet: Lambert et al.  (1995) have obtained HST spectra  
of diffuse interstellar clouds, showing 
the Mulliken (0-0) (2313~\AA) and F-X (0-0) (1342~\AA) bands of C$_{2}$  in 
absorption. CO (e.g., 1509 and 1478 \AA) and excited 
H$_{2}$ (e.g., 1108 and 1092 \AA)
are also prominent absorbers in the
UV, since the column densities in the AGB ejecta are about 
two orders of magnitude larger than for interstellar clouds, the
ultraviolet spectra of post-AGB stars as measured, e.g.\ with 
the HST and IUE, should be dominated by molecular (absorption) bands.

13. Visual: We predict the presence of the CN Violet system and 
    the C$_{3}$ Swing system. Many molecules have been observed 
    in the optical spectra of Comets. Though these are O-rich 
    environments we would expect some of these molecules also
    to be present in C-rich environments. We therefore suggest the
    presence of the Merril-Sandford SiC$_{2}$ bands (4640 and 4977 \AA),
    CH (3130-3150 \AA) and NH (3358 \AA).

14. Infrared: Molecules have many transitions in the infrared. For the stars
in our sample we might expect molecular absorption and emission in the 
infrared of e.g., H$_{2}$ (2-6 $\mu$m), HCN (3.4 $\mu$m), HCN (14.1 $\mu$m), 
C$_{2}$H (27.1 $\mu$m), C$_{2}$H$_{2}$ (13.6-13.8 $\mu$m).

13. Millimeter: Only for the strongest transitions of abundant molecules are
lines in the optical and ultraviolet spectra observed. Many molecular
transitions have been observed in the (sub-)millimeter and radio
for IRC~+10216 (for details see Kawaguchi et al. 1995), 
but very little work has been done to look 
at molecules in these wavelengths for post-AGB stars.
However, to determine accurate abundances and to understand
the circumstellar chemistry of detached dust shells,
observations such as those presented here are of crucial importance.

\subsection{Discussion of individual objects}

\noindent
{\bf IRAS~04296+3429:} Very typical in its circumstellar CN and C$_{2}$
absorption.

\noindent
{\bf IRAS~05113+1347:} Very typical in its circumstellar CN and C$_{2}$
absorption. Our spectra show photospheric CN  bands.

\noindent
{\bf IRAS~05341+0852:} This star has an overabundance of s-process
elements and carbon (Reddy et al. 1997) which clearly suggests that
this is a post-AGB stars. Because of the noisy spectra the
data on C$_{2}$ and CN is not of the same quality as those
for the other stars.

\noindent
{\bf  HD~44179 (The Red Rectangle):} We confirm the presence of CH$^{+}$ (0,0) and (1,0)
in emission at a projected expansion velocity of $3.4\pm2.0$~km s$^{-1}$ with 
$T_{\rm rot}=202\pm16$ K, 
and the  non-detection of C$_{2}$ and CN. CH$^{+}$ in the Red Rectangle
was first detected by Waelkens et al.  (1992) and later identified by
Balm \& Jura (1992)  and Hall et al. (1992) and is so far the
only object which shows CH$^{+}$ in emission. The latter authors estimated the
rotational temperature to be $T_{\rm rot}=120\pm50$ K. 
We prefer our determination of the rotational temperature since many more lines
are included and the error on the equivalent width is significantly smaller.

Although we did not detect CN nor C$_{2}$, the latter molecule is observed
in emission in the reflection lobes of the Red Rectangle (Sarre 1996). Schmidt et al.
(1980) found numerous narrow features in the emission spectrum of the lobes, 
but were not able to make a proper identification. 
Glinsky et al. (1996a) review these unidentified molecular
bands and argue that they are due to phosphorescence of C$_{3}$. 
Glinsky et al. (199b) report on their detection of the spin-forbidden
Cameron bands of CO in the ultraviolet.

The absence of C$_{2}$ and CN absorption could
be related to the low CO/I(60 $\mu$m) ratio (van der Veen et al. 1993). Since
CO is underabundant and the star itself is extremely metal-depleted this might
be the result of the same process: condensation of gas and molecules on 
circumstellar dust grains.
Jura et al.  (1995) detected weak CO millimeter radiation at 
$v_{\ast,\odot}=18.9\pm2.0$~km s$^{-1}$ 
(being the heliocentric stellar velocity derived from CO emission)
with $v_{\rm exp} \approx 6$~km s$^{-1}$. This is close to the CH$^{+}$ 
central 
velocity and  suggests that CH$^{+}$ emission is exactly on the system 
velocity. In the spectra discussed CH$^{+}$ emission lines are not resolved.
New observations (Bakker et al. 1996d) at $R\approx 120,000$ resolve the
line profiles of the stronger lines with a $FWHM\approx 8.5$ km s$^{-1}$
and $FWFM \approx 15$ km s$^{-1}$ (full width  at which the emission falls
below the detection limit). Since the formation of CH$^{+}$ is probably due
to shocks in a circumbinary disk (the emission comes from levels
42000 K above the ground level), the $FWFM$
 is an indication for the turbulent and Kepplerian 
velocity of the line forming region.
The interstellar  CH$^{+}$ abundance  is not very well understood.
Early models (e.g., Elitzur \& Watson 1980) have suggested that CH$^{+}$
might be formed efficiently in shocked regions. Although such models are currently
not favored for the CH$^{+}$ production in interstellar clouds, it could take
place in the complex environment of the Red Rectangle. The fact that no
absorption component is observed suggests that the line-forming
region is not very extended  and is within the ``slit'' of  
the telescope ($\leq 2''$).

It is interesting
to note that the diffuse interstellar bands are in  emission (Sarre 1991),
although we did not find these bands in emission in our spectrum.
Roddier et al. (1995) have spatially resolved the near-infrared emission into
two separate emission peaks with an angular separation of $0.14^{\prime\prime}$. 
Waelkens et al. (1996) 
have clearly demonstrated that it is a  single-lined spectroscopic binary with an 
orbital period of $318\pm3$~days. The latter authors argue that the star is 
not directly observed but only via scattered light on the reflection lobes.

\noindent
{\bf HD~52961:} Like the 
Red Rectangle and HD~213985, the photospheric abundance pattern
shows evidence for selective accretion of circumstellar gas, but 
unlike these two other objects, no CH$^{+}$ has been detected.

\noindent
{\bf  HD~56126:} The C$_{2}$ and CN bands have been studied in detail in 
Paper~I.
It was shown that optical depth effects are important and that 
the excitation of 
C$_{2}$ is a balance between radiative and collisional (de-)excitation. 
HD~56126
has 21 (Kwok et al. 1989) and 30 $\mu$m features (Omont et al. 1995)
and shows many of the s-process elements in its spectrum
(Klochkova 1995). Oudmaijer \& Bakker (1994) suggested a pulsation 
period between 30 and 96 days. More recently L\`{e}bre et al. 
(1996) discuss the possible RV Tauri nature of HD~56126, and they suggest 
the presence of four periods (32.9, 27.3, 11.9, and 7.1 days). We did 
a similar analysis with the data available from L\`{e}bre et al.
and some of our own data and confirm the presence of four significant
minima in the Phase Dispersion diagram on the radial velocities of
HD~56126. However, it seems possible
that there is only one period of $P_{\rm puls}
=12.1$ days and that the other minima are (sub-) harmonics.

\noindent 
{\bf  IRAS~08005-2356:}
Slijkhuis et al.  (1991) have shown that in the optical spectrum all 
absorption lines are blue-shifted up to velocities of 50~km s$^{-1}$. They 
interpreted this as a wind spectrum resulting from a huge post-AGB mass-loss 
rate of about $10^{-6}$~M$_{\odot}$ yr$^{-1}$ with a terminal velocity of about 
400~km s$^{-1}$. The H$\alpha$ profile shows significant changes within a time 
interval of less than two months which suggests a strong asymmetric mass-loss. 
The AGB ejecta emit OH maser emission with a maximum (projected) 
velocity of 50~km s$^{-1}$ (Te Lintel Hekkert 1991), and the infrared 
energy distribution shows the
presence of a cold dust shell ($R \geq 3100$ R$_{\ast}$, 
$T_{\rm dust} \leq 150$~K) and a hot dust shell (or disk) at
$R \geq 34$ R$_{\ast}$ with $T_{\rm dust} \leq 1200$~K.
The overall observational characteristics of IRAS~08005-2356 are very similar 
to those of HD~101584 and suggest that the model by HD~101584 of 
Bakker et al.  (1996b), consisting of a close binary system with an
eccentric orbit also applies to IRAS~08005-2356.
The secondary triggers 
the high-mass loss of the primary and the stellar wind acts as a curtain in 
front of a star much hotter than F5I. This suggests that the optical spectrum 
probably  does not contain information about the star, but only about the wind.

The presence of  C$_{2}$ and C$_{3}$ was also noted
by Hrivnak (1995). Using our high-resolution spectra we are able
to determine the expansion velocity of 50~km s$^{-1}$. 
C$_{2}$ absorption has the
same radial velocity as the blue OH maser emission peak. The fact that the 
OH maser emission and C$_{2}$ are observed at the same velocity is
unexpected since 
the first is a tracer of oxygen rich material (H$_{2}$O dissociation product),
while the latter is an indicator of 
carbon-rich material (C$_{2}$H$_{2}$ dissociation product). 
This suggests that the star has only recently changed from  oxygen to 
carbon-rich without changing its terminal wind velocity, or that there are separate
carbon and oxygen rich  mass-losing stars in the system, or 
that the circumstellar chemistry is very unusual. 
Bakker et al. (1996d) have observed the molecular absorption lines
at a resolution of $R\approx 120,000$ and resolved the molecular
features in two separate component with a line splitting of 
$5.7 \pm 2.0$ km s$^{-1}$. The most likely explanation is that
there are two photodissociation fronts at different velocities: 
one due to the stellar and
one due to  the interstellar radiation field.
We further note 
that there are about twenty chromospheric narrow emission lines 
from neutral and singly ionized metals present in 
our spectrum,  which are double peaked, consistent with
an accretion disk. The presence of an accretion disk, and 
the accretion of material can generate a high velocity bipolar outflow.

We not that this is the only star in our
sample which shows C$_{2}$ and CN but not  the 21 $\mu$m feature.
Very cautiously we propose that this object might have a 21 $\mu$m feature
but that it was excluded from detection so far.

\noindent
{\bf  IRC~+10216:} Much of our knowledge about carbon-rich
circumstellar environments is based on observing and modeling the
carbon-rich AGB star IRC~+10216. Over the years forty different molecular
species have been identified in the radio and (sub)millimeter spectrum
of this star (Lucas 1992). The mass-loss rate and expansion
velocity have been determined from CO observations
by Huggins et al.  (1988)
at $4\pm1\times10^{-5}$~M$_{\odot}$ yr$^{-1}$ and $14\pm1$~km s$^{-1}$.
Circumstellar C$_{2}$ and visible lines of 
CN have not been reported and here we present the first 
detection of C$_{2}$ 
and possibly CN at an expansion velocity of $13.8\pm2.0$~km s$^{-1}$
(Fig.~8). The spectrum is severely blended
and an accurate determination of the continuum level and equivalent widths  
is impossible. 
Taking the local pseudo continuum as the continuum level we have 
determined equivalent widths and derived 
$T_{\rm rot}=43\pm6$~K, $\log N = 14.90\pm0.10$~cm$^{-2}$, and 
$v_{\odot}=-33.3\pm0.7$~km s$^{-1}$,and 
$T_{\rm rot}=26\pm11$~K, $\log N = 14.90 \pm 0.10$~cm$^{-2}$, and 
$v_{\odot}=-32.3\pm0.7$~km s$^{-1}$
for the  C$_{2}$ (2,0) and (3,0) bands, respectively. The theoretical work by 
Cherchneff et al. (1993) predicts
$X_{\rm C_{2}}=4 \times 10^{-6}$ for a distance between 50 and $200 \times 10^{15}$ cm. With a 
mass-loss rate of $3 \times 10^{-5}$ M$_{\odot}$ yr$^{-1}$
and an expansion velocity of 14 km s$^{-1}$ (Huggins et al. 1988)
this gives
a column density of $\log N_{\rm predicted} = 15.30$ cm$^{-2}$. 
Our observations are consistent with
the predictions if we assume that the real continuum is a factor of
three higher than the observed
pseudo continuum. Alternatively, the C$_{2}$ abundance of Cherchneff 
et al. (1993) is a factor three too high.

\noindent
{\bf  HR~4049:}
This truly remarkable object is the most metal-depleted star
known ([Fe/H]$ \sim -4.8$, Waelkens et al. 1991). Bakker et al.
(1996a) have studied the complete optical spectrum at high-resolution
and high signal-to-noise ratio without detecting a single iron peak line.
A strong near-infrared excess has been detected (Lamers et al. 1986)
which is probably due to the presence of a circumbinary disk.
However, unlike the Red Rectangle and HD~213985, no CH$^{+}$ absorption
or emission has been detected in HR~4049, which suggests that the
post-AGB mass-loss, or mass-transfer in the binary system, is less
violent and does not result in the presence of a shocked region 
of circumstellar gas.

\noindent
{\bf HD~161796:} An O-rich supergiant believed to be a post-AGB star.
No molecules have been detected in the optical spectrum of this star.

\noindent
{\bf  IRAS~20000+3239:}
Not much is known about this object except that it exhibits
the unidentified 21~$\mu$m (Kwok et al. 1995) and 
30~$\mu$m (Omont et al. 1995) features. In the sample of
stars with C$_{2}$ and CN absorption lines this is the star
with the highest $N$(CN)/$N$(C$_{2}$) ratio of 11.2, the
hottest CN rotational temperature of $T_{\rm rot}=50\pm12$~K, while it has
a typical expansion velocity of 12.8 km s$^{-1}$. 
The CN column density relative to C$_{2}$ is about twice as high as 
the average.  Our spectra show photospheric CN  bands.
We also note that the CO absorption in the K-band
is the strongest in the sample of stars observed 
(Hrivnak et al. 1994). 

\noindent
{\bf  AFGL~2688 (The Egg Nebula):} This object
is observed as a highly reddened central star surrounded by a torus
(possibly a close binary system) with two bright lobes 
(reflection nebulae) at the equatorial poles of the system.
Three different stellar winds are observed in the CO millimeter
line emission (Young et al. 1992): high (HVW), medium (MVW) and low velocity 
wind (LVW) with expansion velocities of $100\pm10$, 45 and 22.8~km s$^{-1}$,
respectively.  The HVW is bipolar with a de-projected wind velocity of 
360~km s$^{-1}$.
While the  HVW and MVW are post-AGB winds, the LVW is the stellar wind when
the star was on the AGB (now the AGB ejecta). From modeling of the
CO millimeter emission Young et al. found that the AGB ejecta contain
0.7~M$_{\odot}$ fed by an AGB mass-loss rate of  
$\dot M =1.5 \times 10^{-5}$~M$_{\odot}$ yr$^{-1}$. 

C$_{2}$ emission was first reported by  Crampton et al.  (1975) and 
later confirmed by  
Cohen \& Kuhi (1977). From optical spectropolarimetry, Cohen \& Kuhi  
have shown 
that C$_{2}$ is in absorption in the reflected, polarized, 
light of the lobes 
and has an unpolarized emission component, which is attributed to emission 
within the ``slit''. Crampton et al. also report C$_{3}$ absorption.

We have observed the brightest lobe (north lobe) of the Cygnus
Egg Nebula and found C$_{2}$ in absorption with an expansion 
velocity of $17.3\pm2.0$~km s$^{-1}$. This indicates that C$_{2}$ (and CN) is 
formed in the LVW. However, we did not observe any emission, which 
is probably due to the smaller ``slit length'' used, so that only reflected
light from the lobe is observed and no emission from the surrounding gas.
Although reported by Cohen \& Kuhi (1977), 
our spectra do not show the presence 
of the  SiC$_{2}$ Merrill-Sanford band (4977~\AA) in absorption.

We did not detect the  $^{13}$CN Red System  (1,0) band which places
a lower limit of $^{12}$C/$^{13}$C $\geq 19$. This is consistent with the
the isotope ratio of  $^{12}$C/$^{13}$C $\approx 20$ 
found by Wannier \& Sahai (1987) in the slow wind. The fast wind has an isotope
ratio of  $^{12}$C/$^{13}$C $\approx 5$  (Jaminet \& Danchi 1992).
Jaminet \& Danchi found a self absorption feature in the CN millimeter
lines at the same velocity we find the optical absorption lines. 
This opens the possibility to study the CN molecule by both
its optical absorption lines and the millimeter emission lines.

\noindent
{\bf  IRAS~22223+4327:}  Together with HD~235858 these are the only two
sources in our sample which show the CO first overtone in emission. This
is attributed to collisional excitation of the CO molecule in the
circumstellar environment. C$_{2}$ and C$_{3}$ detections are reported 
by Hrivnak (1995). We confirm the presence of C$_{2}$ absorption and add CN to
the list. As a result of the large number of observed molecular 
lines of C$_{2}$ 
the rotational temperature and column density are among the highest in our sample. 

\noindent
{\bf  HD~235858 (IRAS~22272+5435):} This photometric  variable 
has an impressively strong far-infrared excess due to a detached
dust shell. Together with the presence of strong absorption lines from
s-processed elements this is clearly a post-AGB star
(Za{\v c}s et al. 1995). C$_{2}$ and C$_{3}$
absorption in the optical spectrum were first noted by
Hrivnak \& Kwok (1991b), while here we add CN to the list
of detections. Very strong photospheric  CN Red System bands lines 
(Bakker et al. 1996d) and narrow line circumstellar CN  Red System bands
are observed simultaneously (e.g,  7895 to 7905 \AA).
Hrivnak et al.  (1994) reported on the change from
CO first overtone absorption to emission in less than
two months. Since CO first overtone bands are formed in hot gas 
($T_{\rm gas} \sim 1000$ K)
it is not expected that circumstellar C$_{2}$ or CN will follow the variations of CO. 
HD~235858 differs from the other objects showing C$_{2}$ and CN
absorption in that the CN abundance is the lowest in our sample.
Since we have argued that the large spread in observed column densities
of CN is probably due to differences in the UV radiation field 
it seems likely that HD~235858 is exposed to a stronger interstellar
UV radiation field than average.

\noindent
{\bf  HD~213985:} The overall energy distribution is very
much like that of  HR~4049. Differences occur in the ultraviolet because
of different circumstellar extinction laws, and in the far-infrared due to
the presence  of cool AGB ejecta. The infrared energy distribution of 
HD~213985 can be modeled using two  Blackbodies: one component at a distance of 
$22~R_{\ast}$  with $T_{\rm dust}=1250$~K, probably a  circumbinary disk,
and a second component at $220~R_{\ast}$ with $T_{\rm dust}=350$~K, 
the AGB ejecta (Waelkens et al.  1987).  Here we confirm the detection of CH$^{+}$ 
in absorption at an expansion velocity 
of $6.7\pm2.0$~km s$^{-1}$ (Waelkens et al. 1995). If CH$^{+}$ is formed in the
 AGB ejecta at 
a distance of $220~R_{\ast}$  (with $R_{\ast}=50$~R$_{\odot}$) then
 HD~213985 left the AGB only 43~years ago. This would mean that
the effective temperature has increased from 3500~K to 8500~K in 43~years 
giving an average annual increase of 100~K.  
Since this is not observed, CH$^{+}$ cannot
be formed in the AGB ejecta but is probably formed much closer to the star:
a shocked region in a slowly expanding circumbinary disk.

We note that in our spectra of HD~213985 there is evidence for line splitting
very similar to those observed for W Virginis stars (e.g., ST Pup and V29, 
Gonzalez 1993).
Most noticeable the TiII lines at 4549.61 and 4563.77 show not only
a narrow absorption
feature on the red wing, but also the FeII displayed in Fig.~3.

\noindent
{\bf  BD~+39$^o$4926:}
This star is almost a twin of the central star of the Red Rectangle, with the
difference that no infrared excess has been detected for
this object.
Unlike the Red Rectangle and HD~213985, no CH$^{+}$ absorption
or emission has been detected. This combined with the absence
of an infrared excess suggests  that the
post-AGB mass-loss, or mass-transfer in the binary system, is less
violent and does not result in the presence of a shocked region 
of circumstellar gas.

\noindent
{\bf IRAS~23304+6147:} Very typical in its circumstellar CN and C$_{2}$
absorption.

\section{Conclusions}

We have explored a new technique to study the physical and
chemical conditions of the AGB ejecta by looking at molecular
absorption and emission lines in the optical spectra of post-AGB stars.
We find that all stars exhibiting the unidentified 21~$\mu$m feature 
have C$_{2}$ and CN absorption. 
Stars which show C$_{2}$ and CN do not show CH$^{+}$ absorption. The presence
of C$_{2}$ and CN is correlated with the presence  of cold dust ($T_{\rm dust}\leq 300$~K),
while CH$^{+}$ is correlated with the presence of hot dust ($T_{\rm dust}\geq 300$~K).

The expansion velocities determined from the molecular absorption lines
are in very good agreement with the expansion velocities derived from
CO millimeter line emission. This proves that the molecular
absorption lines are formed in the AGB ejecta. The absolute heliocentric 
velocity
of the AGB ejecta in the line-of-sight can be 
determined very accurately thanks to the large number of molecular lines
available in the optical. We find a typical error of $\sim 0.3$~km s$^{-1}$
in our observations.

From the observed equivalent widths (see App.~A, only at CDS) of the absorption lines 
the rotational temperatures and 
column densities can be determined.
We find that the rotational temperature
of C$_{2}$ is significantly higher than that of CN. C$_{2}$ is super-thermally 
excited, whereas CN is sub-thermally excited. This is consistent with 
the fact that C$_{2}$ is a homonuclear molecule (with $I=0$) and 
CN a heteronuclear molecule, while
the primary excitation mechanism of C$_{2}$ is optical 
pumping by the stellar radiation
field. A more detailed analysis of the excitation of these species
can lead to better constraints on the physical parameters
in the AGB ejecta and to an independent determination of the mass-loss
rate. These points will be further investigated in a subsequent paper
(Paper~III in preparation).

Interestingly, we found that the molecular column densities increase with 
expansion velocity. This is interpreted as due to the fact that
carbon-rich dust is accelerated to higher velocities by the
stellar radiation field. The observed column densities
are an indicator of the molecular abundance.
Mass-loss rates are computed which are of the same order of magnitude as those found
from the IR excess and from CO emission lines.
In view of the important assumption made to be able to compute the mass-loss
rate, we stress that these rates should be cited cautiously.

\acknowledgements{The authors want to thank Henny Lamers, Christoffel Waelkens, Ren\'{e} 
Oudmaijer, Hans van Winckel, Guillermo Gonzalez, 
Xander Tielens, John Mathis, and David Lambert
for the stimulating and constructive discussions on this work. 
The significant contributions to this work by Jurien Veenhuis are 
very much appreciated. 
EJB (in the Netherlands) was supported by grant no. 782-371-040 by ASTRON,
which receives funds from the Netherlands Organization for
the Advancement of Pure Research (NWO), 
and (in the USA) in part by the National
Science Foundation (Grant No. AST-9315124).
LBFMW acknowledges financial
support from the Royal Dutch Academy of Arts and Sciences.
EvD is grateful to NWO for support through a PIONIER grant.
This research has made use of the Simbad database, operated at
CDS, Strasbourg, France.}

\newpage
\onecolumn

\appendix

\section{Appendix: Equivalent width of C$_{2}$, CN, and CH$^{+}$ lines}

\noindent
MS 5213 \newline
Circumstellar C$_{2}$, CN, and CH$^{+}$ in the optical spectra of post-AGB stars \newline
Eric J. Bakker, Ewine F. van Dishoeck,  L.B.F.M. Waters, and  Ton Schoenmaker \newline
A\&A main journal, 1997, volume, first page \newline
molecular processes --- circumstellar matter -- stars: AGB and post-AGB --- line: identification \newline
Tables A.1/2/3/4/5 are available in electronic form at the CDS via anonymous
ftp to cdsarc.u-strasbg.fr (130.79.128.5) or via http://cdsweb.u-strasbg.fr/Abstract.html, or
from the authors \newline

\noindent
Description of parameters: \newline
\begin{tabular}{ll}
$B$        & branch identification \\
$J''$      & the rotational quantum number, \\
           & total angular momentum including spin (Herzberg 1950) \\
$N''$      & the rotational quantum number, \\
           & total angular momentum excluding spin (Herzberg 1950) \\
           & for CN: $J''=N'' - 1/2$ (F$_{2}$),  $J''=N'' + 1/2$ (F$_{1}$) \\
$\lambda$  & Laboratory wavelength of transition in air in \AA=1e-10m \\
$f(J'J'')$ & oscillator strength \\
$EW$       & equivalent width in m\AA=1e-13m,\\
           & positive values are absorption lines and \\
           & negative values (only for HD~44179) are emission lines. \\
\end{tabular}

\noindent
Branch identification for  C$_{2}$ $\rm A^{1}\Pi_{u}-X^{1}\Sigma^{+}_{g}$ 
and  CH$^{+}$ $\rm A^{1}\Pi_{u}-X^{1}\Sigma_{g}^{+}$: \newline
\begin{tabular}{lrl}
 $B=$&$-1$ &  P Branch ($\Delta J = -1 = J' - J''$) \\
 $B=$&$0$  &  Q Branch ($\Delta J =  0 = J' - J''$) \\
 $B=$&$1$  &  R Branch ($\Delta J =  1 = J' - J''$) \\
\end{tabular}

\noindent
Branch identification for  CN $\rm A^{2}\Pi-X^{2}\Sigma^{+}$
(after J{\o}rgenson \&  Larsson 1990): \newline
\begin{tabular}{lrl}
 $B=$&$ 1$  &  $\rm R_1     $ \\
 $B=$&$ 2$  &  $\rm Q_1     $ \\
 $B=$&$ 3$  &  $\rm P_1     $ \\
 $B=$&$ 4$  &  $\rm ^QR_{12}$ \\
 $B=$&$ 5$  &  $\rm ^PQ_{12}$ \\
 $B=$&$ 6$  &  $\rm ^OP_{12}$ \\
 $B=$&$ 7$  &  $\rm R_2     $ \\
 $B=$&$ 8$  &  $\rm Q_2     $ \\
 $B=$&$ 9$  &  $\rm P_2     $ \\
 $B=$&$10$  &  $\rm ^SR_{21}$ \\
 $B=$&$11$  &  $\rm ^RQ_{21}$ \\
 $B=$&$12$  &  $\rm ^QP_{21}$ \\
\end{tabular}

\noindent
Code used to identify a star: \newline
\begin{tabular}{ll}
$\ast042$   &  IRAS 04296+3429   \\
$\ast051$   &  IRAS 05113+1347   \\
$\ast053$   &  IRAS 05341+0852   \\
$\ast441$   &  HD 44179          \\
$\ast561$   &  HD 56126          \\
$\ast080$   &  IRAS 08005-2356   \\
$\ast102$   &  IRC +10216        \\
$\ast200$   &  IRAS 20000+3239   \\
$\ast268$   &  AFGL 2688         \\
$\ast222$   &  IRAS 22223+4327   \\
$\ast235$   &  HD 235858         \\
$\ast233$   &  IRAS 23304+6147   \\
$\ast213$   &  HD 213985         \\
\end{tabular}

\noindent
Tables: \newline
\begin{tabular}{ll}
{\bf Table A.1.}  &              C$_2$  $\rm A^{1}\Pi_{u}-X^{1}\Sigma^{+}_{g}$  Phillips   (2,0) band. \\
{\bf Table A.2.}  &              C$_2$  $\rm A^{1}\Pi_{u}-X^{1}\Sigma^{+}_{g}$  Phillips   (3,0) band. \\
{\bf Table A.3.a} &              CN     $\rm A^{2}\Pi-X^{2}\Sigma^{+}$          Red System (2,0) band. \\
{\bf Table A.3.b} &   Continued: CN     $\rm A^{2}\Pi-X^{2}\Sigma^{+}$          Red System (2,0) band. \\
{\bf Table A.4.a} &              CN     $\rm A^{2}\Pi-X^{2}\Sigma^{+}$          Red System (3,0) band. \\
{\bf Table A.4.b} &              CN     $\rm A^{2}\Pi-X^{2}\Sigma^{+}$          Red System (3,0) band. \\
{\bf Table A.5.}  &   Continued: CH$^+$ $\rm A^{1}\Pi-X^{1}\Sigma^{+}$                     (0,0) band. \\
\end{tabular}

\newpage

\begin{tabular}{rrllr}
\multicolumn{5}{l}{{\bf Table A.1.}   C$_2$ $\rm A^{1}\Pi_{u}-X^{1}\Sigma^{+}_{g}$  Phillips (2,0) band.} \\
\hline
\hline
$B$&$J''$&$\lambda$ (air) [\AA]&$f(J'J'')$&$EW$ [m\AA] \\
\cline{5-5}
   &   &                     &   & $\ast102$   \\
\hline
  &   &         &         &       \\
 1&  6& 8750.850& 4.43E-04&  44.3 \\ 
 1&  8& 8751.490& 4.24E-04&       \\
 1&  4& 8751.687& 4.80E-04& 117.5 \\
 1& 10& 8753.581& 4.11E-04&       \\
 1&  2& 8753.949& 5.76E-04&  77.1 \\
 1& 12& 8757.130& 4.04E-04&       \\
 1&  0& 8757.686& 1.44E-03& 117.5 \\
 0&  2& 8761.197& 7.20E-04& 125.0 \\
 1& 14& 8762.147& 3.98E-04&       \\
 0&  4& 8763.754& 7.20E-04& 118.2 \\
-1&  2& 8766.031& 1.44E-04&  53.9 \\
 0&  6& 8767.762& 7.19E-04&  32.7 \\
 1& 16& 8768.631& 3.92E-04&       \\
 0&  8& 8773.223& 7.19E-04&       \\
-1&  4& 8773.430& 2.40E-04&  29.6 \\
 1& 18& 8776.611& 3.88E-04&       \\
 0& 10& 8780.144& 7.18E-04&       \\
-1&  6& 8782.311& 2.76E-04&       \\
 1& 20& 8786.050& 3.85E-04&       \\
 0& 12& 8788.561& 7.17E-04&       \\
-1&  8& 8792.652& 2.96E-04&       \\
 1& 22& 8796.979& 3.82E-04&       \\
 0& 14& 8798.462& 7.17E-04&       \\
-1& 10& 8804.502& 3.07E-04&       \\
 1& 24& 8809.410& 3.80E-04&       \\
 0& 16& 8809.844& 7.16E-04&       \\
-1& 12& 8817.830& 3.15E-04&       \\
 0& 18& 8822.728& 7.15E-04&       \\
 1& 26& 8823.379& 3.78E-04&       \\
-1& 14& 8832.682& 3.20E-04&       \\
 0& 20& 8837.122& 7.14E-04&       \\
 1& 28& 8838.846& 3.76E-04&       \\
-1& 16& 8849.075& 3.24E-04&       \\
 0& 22& 8853.044& 7.12E-04&       \\
 1& 30& 8855.876& 3.73E-04&       \\
-1& 18& 8866.997& 3.27E-04&       \\
 0& 24& 8870.519& 7.11E-04&       \\
 1& 32& 8874.466& 3.72E-04&       \\
-1& 20& 8886.487& 3.29E-04&       \\
 0& 26& 8889.535& 7.10E-04&       \\
 1& 34& 8894.623& 3.70E-04&       \\
-1& 22& 8907.545& 3.30E-04&       \\
 0& 28& 8910.135& 7.08E-04&       \\
 1& 36& 8916.400& 3.68E-04&       \\
-1& 24& 8930.172& 3.31E-04&       \\
 0& 30& 8932.333& 7.06E-04&       \\
-1& 26& 8954.440& 3.32E-04&       \\
 0& 32& 8956.139& 7.04E-04&       \\
-1& 28& 8980.336& 3.33E-04&       \\
 0& 34& 8981.586& 7.02E-04&       \\
-1& 30& 9007.897& 3.33E-04&       \\
 0& 36& 9008.705& 7.00E-04&       \\
-1& 32& 9037.150& 3.33E-04&       \\
-1& 34& 9068.086& 3.33E-04&       \\
-1& 36& 9100.823& 3.32E-04&       \\
  &   &         &         &       \\
\hline
\hline
\end{tabular}

\newpage

\begin{tabular}{rrllrrrrrrrrrrr}
\multicolumn{14}{l}{{\bf Table A.2.}  C$_2$  $\rm A^{1}\Pi_{u}-X^{1}\Sigma^{+}_{g}$  
Phillips (3,0) band.} \\
\hline
\hline
$B$&$J''$&$\lambda$ (air) [\AA]&$f(J'J'')$&
                           \multicolumn{11}{c}{$EW$ [m\AA]} \\
\cline{5-15}
   &   &                     &   &
$\ast042$& $\ast051$& $\ast053$& $\ast561$&$\ast080$&$\ast102$& $\ast200$ & $\ast268$&
$\ast222$ & $\ast235$& $\ast233$ \\
\hline
  &   &         &         &      &      &      &      &      &      &      &      &      &      &      \\
 1&  6& 7714.578& 2.006-04&  34.9&  60.5&  16.8&  28.0&  85.0&      &  33.4&  43.3&  52.1&  40.0&  70.7\\
 1&  4& 7714.947& 2.23E-04&  47.9&  34.0&  27.4&  30.7&  95.0&      &  33.6&  76.6&  61.9&  17.0&  73.2\\
 1&  8& 7715.418& 1.96E-04&  22.1&  28.0&  11.6&  24.0&  40.3&      &  21.9&  25.7&  47.4&  12.0&  51.5\\
 1&  2& 7716.531& 2.67E-04&  53.9&  23.8&      &  29.5& 100.0&      &  57.2& 133.9&  66.9&  36.7&  62.2\\
 1& 10& 7717.473& 1.91E-04&      &      &      &  14.2&  31.9&      &  25.1&      &  29.1&      &  35.0\\
 1&  0& 7719.331& 6.67E-04&  58.3&  27.3&  34.2&  30.5&  73.4&  72.0&  55.2& 152.7&  58.9&  59.5&  56.4\\
 1& 12& 7720.751& 1.87E-04&      &  13.4&      &  23.1&  24.7&      &  25.1&      &  21.6&      &  23.3\\
 0&  2& 7722.096& 3.34E-04&  56.2&  22.1&  44.9&  34.7&  91.7&  44.0&  67.6& 152.4&  75.2&  47.5&  78.2\\
 0&  4& 7724.222& 3.34E-04&  61.6&  26.1&  30.4&      &  94.0&  38.0&  63.6&  98.3&  82.9&  59.9&  77.2\\
 1& 14& 7725.243& 1.84E-04&      &      &      &      &  12.8&      &  19.1&      &  18.0&      &  12.4\\
-1&  2& 7725.822& 6.65E-05&  30.9&  12.4&   9.6&  16.4&  55.8&      &  33.8&  70.1&  28.9&  22.0&  38.7\\
 0&  6& 7727.560& 3.34E-04&  41.5&      &  49.0&      &  84.8&      &  25.8&      &  78.3&  20.0&  71.7\\
 1& 16& 7730.966& 1.82E-04&      &   4.3&      &  11.4&      &      &   9.7&      &  14.0&      &   8.3\\
-1&  4& 7731.666& 1.11E-04&  14.0&   9.1&  26.2&  29.2&  70.1&      &  39.5&  44.9&  39.2&  45.0&  50.6\\
 0&  8& 7732.120& 3.33E-04&  22.7&  15.5&  18.2&  29.6&  61.3&      &  40.5&      &  64.9&  33.0&  82.0\\
 0& 10& 7737.908& 3.33E-04&  18.9&  18.3&      &      &  44.7&      &  32.2&      &  55.0&  65.6&  53.4\\
 1& 18& 7737.908& 1.80E-04&      &      &      &      &      &      &      &      &      &      &      \\
-1&  6& 7738.740& 1.28E-04&  10.3&  11.9&      &  20.6&  51.8&      &  32.6&      &  36.1&  33.5&  51.5\\
 0& 12& 7744.903& 3.33E-04&      &      &      &      &  27.3&      &      &      &      &      &      \\
 1& 20& 7746.098& 1.78E-04&      &      &      &      &      &      &      &      &      &      &      \\
-1&  8& 7747.040& 1.37E-04&      &      &      &      &  47.5&      &      &      &      &      &      \\
 0& 14& 7753.144& 3.32E-04&      &      &      &      &  16.1&      &      &      &      &      &      \\
 1& 22& 7755.532& 1.77E-04&      &      &      &      &      &      &      &      &      &      &      \\
-1& 10& 7756.585& 1.42E-04&      &      &      &      &  16.0&      &      &      &      &      &      \\
 0& 16& 7762.626& 3.31E-04&      &      &      &      &  11.0&      &      &      &      &      &      \\
 1& 24& 7766.226& 1.76E-04&      &      &      &      &      &      &      &      &      &      &      \\
-1& 12& 7767.372& 1.46E-04&      &      &      &  13.7&  12.9&      &      &      &      &      &      \\
 0& 18& 7773.358& 3.31E-04&   1.0&      &      &      &      &      &  14.3&      &  12.0&      &   7.3\\
 1& 26& 7778.178& 1.75E-04&      &      &      &      &      &      &      &      &      &      &      \\
-1& 14& 7779.431& 1.48E-04&      &      &      &  14.0&      &      &      &      &  13.7&      &  12.8\\
 0& 20& 7785.350& 3.31E-04&      &      &      &      &      &      &      &      &  17.2&      &  15.7\\
 1& 28& 7791.405& 1.74E-04&      &      &      &      &      &      &      &      &      &      &      \\
-1& 16& 7792.741& 1.50E-04&      &      &      &      &      &      &      &      &  12.0&      &  14.9\\
 0& 22& 7798.613& 3.30E-04&      &      &      &      &      &      &      &      &  10.1&      &  11.2\\
 1& 30& 7805.908& 1.73E-04&      &      &      &      &      &      &      &      &      &      &      \\
-1& 18& 7807.322& 1.52E-04&      &      &      &      &      &      &      &      &   9.8&      &   9.6\\
 0& 24& 7813.138& 3.30E-04&      &      &      &      &      &      &      &      &  15.6&      &   4.5\\
 1& 32& 7821.720& 1.72E-04&      &      &      &      &      &      &      &      &      &      &      \\
-1& 20& 7823.201& 1.53E-04&      &      &      &      &      &      &      &      &   4.7&      &      \\
 0& 26& 7828.979& 3.29E-04&      &      &      &      &      &      &      &      &   8.2&      &      \\
 1& 34& 7838.850& 1.71E-04&      &      &      &      &      &      &      &      &      &      &      \\
-1& 22& 7840.387& 1.53E-04&      &      &      &      &      &      &      &      &      &      &      \\
 0& 28& 7846.122& 3.28E-04&      &      &      &      &      &      &      &      &  10.2&      &      \\
 1& 36& 7857.314& 1.70E-04&      &      &      &      &      &      &      &      &      &      &      \\
-1& 24& 7858.889& 1.53E-04&      &      &      &      &      &      &      &      &      &      &      \\
 0& 30& 7864.595& 3.27E-04&      &      &      &      &      &      &      &      &      &      &      \\
-1& 26& 7878.721& 1.54E-04&      &      &      &      &      &      &      &      &      &      &      \\
 0& 32& 7884.394& 3.27E-04&      &      &      &      &      &      &      &      &      &      &      \\
-1& 28& 7899.907& 1.54E-04&      &      &      &      &      &      &      &      &      &      &      \\
 0& 34& 7905.561& 3.26E-04&      &      &      &      &      &      &      &      &      &      &      \\
-1& 30& 7922.464& 1.54E-04&      &      &      &      &      &      &      &      &      &      &      \\
 0& 36& 7928.100& 3.25E-04&      &      &      &      &      &      &      &      &      &      &      \\
-1& 32& 7946.406& 1.54E-04&      &      &      &      &      &      &      &      &      &      &      \\
-1& 34& 7971.783& 1.54E-04&      &      &      &      &      &      &      &      &      &      &      \\
-1& 36& 7998.571& 1.54E-04&      &      &      &      &      &      &      &      &      &      &      \\
  &   &         &         &      &      &      &      &      &      &      &      &      &      &      \\
\hline
\hline
\end{tabular}

\newpage

\begin{tabular}{rrrllrrr}
\multicolumn{8}{l}{{\bf Table A.3.a}   CN $\rm A^{2}\Pi-X^{2}\Sigma^{+}$  Red System (2,0) 
band.} \\
\hline
\hline
$B$&$J''$&$N''$&$\lambda$ (air) [\AA]&$f(J'J'')$&
                                 \multicolumn{3}{c}{$EW$ [m\AA]} \\
\cline{6-8}
   &   &   &                     &   &
                                 $\ast080$&$\ast102$&$\ast235$ \\
\hline
  &    &   &         &         &      &      &       \\
 6& 8.5&  9& 7952.261& 4.49E-05&      &      &       \\
 6& 7.5&  8& 7946.181& 4.65E-05&      &      &       \\
 6& 6.5&  7& 7940.433& 4.71E-05&      &      &       \\
 6& 5.5&  6& 7935.020& 4.69E-05&      &      &       \\
 3& 9.5&  9& 7934.855& 1.36E-04&      &      &       \\
 5& 8.5&  9& 7934.812& 1.22E-04&      &      &       \\
 3& 8.5&  8& 7930.849& 1.27E-04&      &      &       \\
 5& 7.5&  8& 7930.811& 1.30E-04&      &      &       \\
 6& 4.5&  5& 7929.943& 4.46E-05&      &      &       \\
 3& 7.5&  7& 7927.167& 1.18E-04&      &      &       \\
 5& 6.5&  7& 7927.134& 1.40E-04&      &      &       \\
 6& 3.5&  4& 7925.204& 3.84E-05&      &      &       \\
 3& 6.5&  6& 7923.813& 1.08E-04&      &      &       \\
 5& 5.5&  6& 7923.784& 1.51E-04&      &      &       \\
 6& 2.5&  3& 7920.804& 2.56E-05&      &      &       \\
 3& 5.5&  5& 7920.786& 9.53E-05&      &      &       \\
 5& 4.5&  5& 7920.762& 1.61E-04&      &      &       \\
 3& 4.5&  4& 7918.091& 8.04E-05&      &      &       \\
 5& 3.5&  4& 7918.071& 1.70E-04&  17.3&      &       \\
 3& 3.5&  3& 7915.729& 6.22E-05&      &      &       \\
 5& 2.5&  3& 7915.714& 1.75E-04&  43.9&      &       \\
 2& 9.5&  9& 7915.407& 3.49E-04&      &      &       \\
 4& 8.5&  9& 7915.364& 7.82E-05&      &      &       \\
 3& 2.5&  2& 7913.704& 3.64E-05&      &      &       \\
 5& 1.5&  2& 7913.692& 1.64E-04&      &      &       \\
 2& 8.5&  8& 7913.473& 3.40E-04&      &      &       \\
 4& 7.5&  8& 7913.435& 8.68E-05&      &      &       \\
 9& 8.5&  9& 7913.336& 1.49E-04&      &      &       \\
 2& 7.5&  7& 7911.856& 3.29E-04&      &      &       \\
 4& 6.5&  7& 7911.822& 9.73E-05&      &      &       \\
 2& 6.5&  6& 7910.559& 3.18E-04&      &      &       \\
 4& 5.5&  6& 7910.529& 1.10E-04&      &      &       \\
 2& 5.5&  5& 7909.584& 3.05E-04&      &      &       \\
 4& 4.5&  5& 7909.559& 1.26E-04&      &      &       \\
 2& 1.5&  1& 7908.966& 2.08E-04&      &      &       \\
 4& 0.5&  1& 7908.959& 4.37E-04&  65.1&  39.4&       \\
 2& 4.5&  4& 7908.934& 2.91E-04&      &      &       \\
 4& 3.5&  4& 7908.914& 1.48E-04&      &      &       \\
 2& 2.5&  2& 7908.622& 2.50E-04& 109.1&  52.2&       \\
 2& 3.5&  3& 7908.613& 2.73E-04&      &      &       \\
 4& 1.5&  2& 7908.611& 2.45E-04&      &      &       \\
 4& 2.5&  3& 7908.597& 1.81E-04&      &      &       \\
 9& 7.5&  8& 7908.172& 1.43E-04&      &      &       \\
 1& 0.5&  0& 7906.598& 4.93E-04&  36.1&  37.9&  36.9 \\
 1& 1.5&  1& 7903.892& 3.12E-04&  47.2&  71.5&  18.2 \\
 9& 6.5&  7& 7903.237& 1.37E-04&      &      &       \\
 1& 2.5&  2& 7901.520& 2.59E-04&  30.9&      &  23.1 \\
 1& 3.5&  3& 7899.481& 2.37E-04&  33.5&      &       \\
 9& 5.5&  6& 7898.529& 1.29E-04&      &      &       \\
 1& 4.5&  4& 7897.771& 2.26E-04&  38.5&      &       \\
 1& 5.5&  5& 7896.386& 2.21E-04&  15.7&      &       \\
 1& 6.5&  6& 7895.326& 2.18E-04&      &      &       \\
\hline
\end{tabular}

\newpage

\begin{tabular}{rrrllrrr}
\multicolumn{8}{l}{{\bf Table A.3.b}  Continued: CN $\rm A^{2}\Pi-X^{2}\Sigma^{+}$ 
Red System (2,0) band.} \\
\hline
$B$&$J''$&$N''$&$\lambda$ (air) [\AA]&$f(J'J'')$&
                                 \multicolumn{3}{c}{$EW$ [m\AA]} \\
\cline{6-8}
   &   &   &                     &   &
                                 $\ast080$&$\ast102$&$\ast235$ \\
\hline
12& 9.5&  9& 7895.174& 6.22E-05&      &      &       \\
 8& 8.5&  9& 7895.131& 3.42E-04&      &      &       \\
 1& 7.5&  7& 7894.583& 2.17E-04&      &      &       \\
 1& 8.5&  8& 7894.158& 2.17E-04&      &      &       \\
 1& 9.5&  9& 7894.045& 2.17E-04&      &      &       \\
 9& 4.5&  5& 7894.043& 1.21E-04&      &      &       \\
12& 8.5&  8& 7892.153& 6.68E-05&      &      &       \\
 8& 7.5&  8& 7892.116& 3.32E-04&      &      &       \\
 9& 3.5&  4& 7889.776& 1.11E-04&      &      &       \\
12& 7.5&  7& 7889.360& 7.14E-05&      &      &       \\
 8& 6.5&  7& 7889.327& 3.22E-04&      &      &       \\
12& 6.5&  6& 7886.792& 7.60E-05&      &      &       \\
 8& 5.5&  6& 7886.763& 3.11E-04&      &      &       \\
 9& 2.5&  3& 7885.725& 9.84E-05&      &      &       \\
12& 5.5&  5& 7884.445& 8.06E-05&      &      &       \\
 8& 4.5&  5& 7884.420& 2.99E-04&      &      &       \\
12& 4.5&  4& 7882.315& 8.46E-05&      &      &       \\
 8& 3.5&  4& 7882.295& 2.88E-04&  23.5&  18.7&       \\
 9& 1.5&  2& 7881.889& 7.79E-05&      &      &       \\
12& 3.5&  3& 7880.400& 8.78E-05&      &      &       \\
 8& 2.5&  3& 7880.384& 2.77E-04&  38.2&      &       \\
12& 2.5&  2& 7878.697& 8.76E-05&      &      &       \\
 8& 1.5&  2& 7878.686& 2.70E-04&  54.1&      &       \\
12& 1.5&  1& 7877.205& 7.79E-05&      &      &       \\
 8& 0.5&  1& 7877.198& 3.09E-04&  28.8&      &       \\
11& 9.5&  9& 7874.878& 1.14E-04&      &      &       \\
11& 0.5&  0& 7874.852& 3.09E-04&  37.1&      &       \\
 7& 8.5&  9& 7874.835& 1.93E-04&      &      &       \\
11& 1.5&  1& 7873.992& 2.26E-04&  51.1&      &       \\
11& 8.5&  8& 7873.987& 1.23E-04&      &      &       \\
 7& 0.5&  1& 7873.985& 1.84E-04&      &      &       \\
 7& 7.5&  8& 7873.949& 1.90E-04&      &      &       \\
11& 2.5&  2& 7873.343& 2.01E-04&  41.3&      &       \\
 7& 1.5&  2& 7873.332& 1.73E-04&      &      &       \\
11& 7.5&  7& 7873.325& 1.34E-04&      &      &       \\
 7& 6.5&  7& 7873.292& 1.86E-04&      &      &       \\
11& 3.5&  3& 7872.905& 1.84E-04&      &      &       \\
 7& 2.5&  3& 7872.889& 1.73E-04&      &      &       \\
11& 6.5&  6& 7872.889& 1.45E-04&      &      &       \\
 7& 5.5&  6& 7872.860& 1.82E-04&      &      &       \\
11& 4.5&  4& 7872.682& 1.70E-04&      &      &       \\
11& 5.5&  5& 7872.676& 1.57E-04&      &      &       \\
 7& 3.5&  4& 7872.662& 1.75E-04&      &      &       \\
 7& 4.5&  5& 7872.652& 1.78E-04&      &      &       \\
10& 0.5&  0& 7871.654& 1.28E-04&      &      &       \\
10& 1.5&  1& 7868.667& 1.07E-04&  20.7&      &       \\
10& 2.5&  2& 7865.892& 9.53E-05&  21.2&      &       \\
10& 3.5&  3& 7863.333& 8.60E-05&      &      &       \\
  &    &   &         &         &      &      &       \\
\hline
\hline
\end{tabular}

\newpage

\begin{tabular}{rrrllrrrrrrrr}
\multicolumn{12}{l}{{\bf Table A.4.a}  CN $\rm A^{2}\Pi-X^{2}\Sigma^{+}$ Red System (3,0) band.} \\
\hline
\hline
$B$&$J''$&$N''$&$\lambda$ (air) [\AA]&$f(J'J'')$&
                                 \multicolumn{8}{c}{$EW$ [m\AA]} \\
\cline{6-13}
   &   &   &                     &   &
                                 $\ast042$&$\ast051$&$\ast053$&$\ast561$&$\ast200$&
                                 $\ast268$&$\ast222$ &$\ast233$ \\
\hline
  &    &   &         &         &      &      &      &      &      &      &      &      \\
 6& 8.5&  9& 6987.583& 1.98E-05&      &      &      &      &      &      &      &      \\
 6& 7.5&  8& 6982.767& 2.04E-05&      &      &      &      &      &      &      &      \\
 6& 6.5&  7& 6978.223& 2.08E-05&      &      &      &      &      &      &      &      \\
 3& 9.5&  9& 6974.278& 5.93E-05&      &      &      &      &      &      &      &      \\
 5& 8.5&  9& 6974.244& 5.33E-05&      &      &      &      &      &      &      &      \\
 6& 5.5&  6& 6973.953& 2.05E-05&      &      &      &      &      &      &      &      \\
 3& 8.5&  8& 6971.045& 5.57E-05&      &      &      &      &      &      &      &      \\
 5& 7.5&  8& 6971.016& 5.74E-05&      &      &      &      &      &      &      &      \\
 6& 4.5&  5& 6969.958& 1.96E-05&      &      &      &      &      &      &      &      \\
 3& 7.5&  7& 6968.079& 5.18E-05&      &      &      &      &      &      &      &      \\
 5& 6.5&  7& 6968.053& 6.18E-05&      &      &      &      &      &      &      &      \\
 6& 3.5&  4& 6966.241& 1.68E-05&      &      &      &      &      &      &      &   6.5\\
 3& 6.5&  6& 6965.383& 4.71E-05&      &      &      &      &      &      &      &      \\
 5& 5.5&  6& 6965.360& 6.62E-05&      &      &      &      &      &      &      &      \\
 3& 5.5&  5& 6962.956& 4.18E-05&      &      &      &      &      &      &      &      \\
 5& 4.5&  5& 6962.937& 7.08E-05&      &      &      &      &  79.7&      &      &      \\
 6& 2.5&  3& 6962.801& 1.12E-05&  11.5&      &      &      &      &      &      &      \\
 3& 4.5&  4& 6960.801& 3.53E-05&      &      &      &      &      &      &      &      \\
 5& 3.5&  4& 6960.785& 7.47E-05&   8.2&      &      &      &  70.3&      &  73.2&  53.2\\
 2& 9.5&  9& 6959.402& 1.53E-04&      &      &      &      &      &      &      &      \\
 4& 8.5&  9& 6959.369& 3.45E-05&      &      &      &      &      &      &      &      \\
 3& 3.5&  3& 6958.920& 2.70E-05&      &      &      &      &      &      &      &      \\
 5& 2.5&  3& 6958.908& 7.68E-05&  11.5&  25.6&  46.8&  12.8&  76.6&  19.2&  72.9&  43.2\\
 2& 8.5&  8& 6957.753& 1.48E-04&      &      &      &      &      &      &      &      \\
 4& 7.5&  8& 6957.724& 3.83E-05&      &      &      &      &      &      &      &      \\
 9& 8.5&  9& 6957.622& 6.54E-05&      &      &      &      &      &      &      &      \\
 3& 2.5&  2& 6957.314& 1.60E-05&      &      &      &      &      &      &      &      \\
 5& 1.5&  2& 6957.305& 7.19E-05&  19.3&  42.2&      &   7.9&  58.6&  30.7&  67.4&      \\
 2& 7.5&  7& 6956.367& 1.44E-04&      &      &      &      &      &      &      &      \\
 4& 6.5&  7& 6956.340& 4.28E-05&      &      &      &      &      &      &      &      \\
 2& 6.5&  6& 6955.243& 1.39E-04&      &      &      &      &      &      &      &      \\
 4& 5.5&  6& 6955.220& 4.84E-05&      &      &      &      &      &      &      &      \\
 2& 5.5&  5& 6954.385& 1.33E-04&      &      &      &      &      &      &      &      \\
 4& 4.5&  5& 6954.366& 5.55E-05&      &      &      &      &      &      &      &      \\
 2& 4.5&  4& 6953.795& 1.27E-04&      &      &      &      &      &      &      &      \\
 4& 3.5&  4& 6953.779& 6.50E-05&      &      &      &      &      &      &      &      \\
 2& 1.5&  1& 6953.652& 9.13E-05&      &      &      &      &      &      &      &      \\
 4& 0.5&  1& 6953.646& 1.92E-04&      &  58.2&      &      &      &      &      &      \\
 9& 7.5&  8& 6953.498& 6.27E-05&      &      &      &      &      &      &      &      \\
 2& 3.5&  3& 6953.475& 1.20E-04&      &      &      &      &      &      &      &      \\
 4& 2.5&  3& 6953.462& 7.99E-05&      &      &      &      &      &      &      &      \\
 2& 2.5&  2& 6953.427& 1.10E-04&      &      &      &      &      &      &      &      \\
 4& 1.5&  2& 6953.418& 1.08E-04&      &      &      &      &      &      &      &      \\
 1& 0.5&  0& 6951.821& 2.15E-04&  70.9& 106.8&      &  34.1& 122.9& 104.2& 145.2& 110.8\\
 1& 1.5&  1& 6949.770& 1.36E-04&  15.2&      &  34.1&  24.0& 101.0& 114.4& 128.9&  55.7\\
 9& 6.5&  7& 6949.568& 5.98E-05&      &      &      &      &      &      &      &      \\
 1& 2.5&  2& 6947.992& 1.14E-04&  20.9&      &  42.0&  14.3& 152.5&  66.9& 115.7&  44.3\\
 1& 3.5&  3& 6946.486& 1.04E-04&      &      &      &      &  96.4&      &  87.1&  37.7\\
 9& 5.5&  6& 6945.831& 5.66E-05&      &      &      &      &      &      &      &      \\
 1& 4.5&  4& 6945.251& 9.91E-05&      &      &      &      &      &      &      &      \\
 1& 5.5&  5& 6944.285& 9.66E-05&      &      &  67.0&      &      &      &      &      \\
12& 9.5&  9& 6943.727& 2.74E-05&      &      &      &      &      &      &      &      \\
\hline
\end{tabular}

\newpage

\begin{tabular}{rrrllrrrrrrrr}
\multicolumn{12}{l}{{\bf Table A.4.b} Continued: CN $\rm A^{2}\Pi-X^{2}\Sigma^{+}$ 
Red System (3,0) band.} \\
\hline
$B$&$J''$&$N''$&$\lambda$ (air) [\AA]&$f(J'J'')$&
                                 \multicolumn{8}{c}{$EW$ [m\AA]} \\
\cline{6-13}
   &   &   &                     &   &
                                 $\ast042$&$\ast051$&$\ast053$&$\ast561$&$\ast200$&
                                 $\ast268$&$\ast222$ &$\ast233$ \\
\hline
 8& 8.5&  9& 6943.693& 1.50E-04&      &      &      &      &      &      &      &      \\
 1& 6.5&  6& 6943.584& 9.53E-05&      &      &      &      &      &      &      &      \\
 1& 7.5&  7& 6943.147& 9.48E-05&      &      &      &      &      &      &      &      \\
 1& 9.5&  9& 6943.052& 9.49E-05&      &      &      &      &      &      &      &      \\
 1& 8.5&  8& 6942.970& 9.48E-05&      &      &      &      &      &      &      &      \\
 9& 4.5&  5& 6942.282& 5.30E-05&      &      &      &      &      &      &      &      \\
12& 8.5&  8& 6941.241& 2.94E-05&      &      &      &      &      &      &      &      \\
 8& 7.5&  8& 6941.211& 1.45E-04&      &      &      &      &      &      &      &      \\
12& 7.5&  7& 6938.950& 3.14E-05&      &      &      &      &      &      &      &      \\
 8& 6.5&  7& 6938.923& 1.41E-04&      &      &      &      &      &      &      &      \\
 9& 3.5&  4& 6938.920& 4.87E-05&   9.7&      &      &      &  68.3&      &      &  52.3\\
12& 6.5&  6& 6936.850& 3.35E-05&      &      &      &      &      &      &      &      \\
 8& 5.5&  6& 6936.827& 1.36E-04&      &      &      &      &      &      &      &      \\
 9& 2.5&  3& 6935.742& 4.32E-05&      &      &      &      &      &  34.6&  37.9&  35.0\\
12& 5.5&  5& 6934.936& 3.55E-05&      &      &      &      &      &      &      &      \\
 8& 4.5&  5& 6934.917& 1.31E-04&      &      &      &      &  53.9&      &      &      \\
12& 4.5&  4& 6933.209& 3.71E-05&   8.6&      &      &      &      &      &      &      \\
 8& 3.5&  4& 6933.194& 1.26E-04&      &  14.0&      &      &  76.0&      &  54.8&  43.5\\
 9& 1.5&  2& 6932.748& 3.42E-05&      &      &      &      &      &      &  33.4&      \\
12& 3.5&  3& 6931.667& 3.85E-05&      &      &      &      &      &      &      &      \\
 8& 2.5&  3& 6931.655& 1.21E-04&  14.2&      &      &      &  97.4&      &  84.6&  41.5\\
12& 2.5&  2& 6930.305& 3.84E-05&      &      &      &      &      &      &      &      \\
 8& 1.5&  2& 6930.297& 1.18E-04&  43.5&      &  11.1&  12.9&  93.7&  87.8& 119.8&  56.7\\
12& 1.5&  1& 6929.124& 3.42E-05&      &      &      &      &      &      &      &      \\
 8& 0.5&  1& 6929.119& 1.36E-04&      &  56.9&      &  18.9& 111.9&  80.8& 100.4&  37.4\\
11& 9.5&  9& 6928.188& 5.02E-05&      &      &      &      &      &      &      &      \\
 7& 8.5&  9& 6928.155& 8.46E-05&      &      &      &      &      &      &      &      \\
11& 8.5&  8& 6927.333& 5.44E-05&      &      &      &      &      &      &      &      \\
 7& 7.5&  8& 6927.303& 8.30E-05&      &      &      &      &      &      &      &      \\
11& 0.5&  0& 6927.302& 1.36E-04&  35.5&      &      &  27.1&  85.7& 107.1&  85.1&      \\
11& 7.5&  7& 6926.673& 5.89E-05&      &      &      &      &      &      &      &      \\
11& 1.5&  1& 6926.665& 9.90E-05&  37.9&      &      &      &  95.0&  32.0&      &      \\
 7& 0.5&  1& 6926.659& 8.06E-05&      &      &  92.9&      &      &      &      &      \\
 7& 6.5&  7& 6926.646& 8.14E-05&      &      &      &      &      &      &      &      \\
11& 2.5&  2& 6926.206& 8.84E-05&      &      &  11.3&  17.2&      &      &      &      \\
11& 6.5&  6& 6926.205& 6.37E-05&      &      &      &      &      &      &      &      \\
 7& 1.5&  2& 6926.197& 7.60E-05&  42.0&      &      &      & 144.5&      &      &      \\
 7& 5.5&  6& 6926.182& 7.97E-05&      &      &      &      &      &      &      &      \\
11& 3.5&  3& 6925.929& 8.11E-05&      &      &  81.4&      &      &      &      &      \\
11& 5.5&  5& 6925.926& 6.90E-05&      &      &      &      &      &      &      &      \\
 7& 2.5&  3& 6925.917& 7.58E-05&  38.8&      &      &      & 158.9&      &      &  72.2\\
 7& 4.5&  5& 6925.907& 7.81E-05&      &      &      &      &      &      &      &      \\
11& 4.5&  4& 6925.835& 7.46E-05&      &      &      &      &      &      &      &      \\
 7& 3.5&  4& 6925.819& 7.67E-05&      &      &      &      &      &      &      &      \\
10& 0.5&  0& 6924.855& 5.61E-05&  58.5&      &  62.0&      &  44.2&  92.7&  96.8&      \\
10& 1.5&  1& 6922.587& 4.74E-05&  26.3&      &      &      &  65.4&  81.5&  71.1&  63.0\\
10& 2.5&  2& 6920.501& 4.18E-05&      &      &      &      &  49.0&      &      &      \\
10& 3.5&  3& 6918.598& 3.77E-05&      &      &      &      &     0&      &      &      \\
  &    &   &         &         &      &      &      &      &      &      &      &      \\
\hline
\hline
\end{tabular}

\newpage

\begin{tabular}{rrllrr}
\multicolumn{6}{l}{{\bf Table A.5.} CH$^+$  $\rm A^{1}\Pi-X^{1}\Sigma^{+}$ (0,0) band.} \\
\hline
\hline
$B$&$J''$&$\lambda$ (air) [\AA]&$f(J'J'')$&\multicolumn{2}{c}{$EW$ [m\AA]} \\
\cline{5-6}
   &   &                     &    &
                              $\ast441$&$\ast213$\\
\hline
  &   &         &         &      &      \\
 1&  4& 4225.276& 1.81E-03&  -9.0&   4.0\\
 1&  3& 4225.701& 1.94E-03& -14.3&  10.8\\
 1&  5& 4225.807& 1.73E-03&      &      \\
 1&  2& 4227.064& 2.18E-03& -12.7&  30.0\\
 1&  6& 4227.318& 1.67E-03&      &      \\
 1&  1& 4229.351& 2.72E-03& -25.2&  18.8\\
 1&  7& 4229.837& 1.63E-03&      &      \\
 1&  0& 4232.552& 5.45E-03& -18.0&  37.6\\
 1&  8& 4233.398& 1.60E-03&      &      \\
 1&  9& 4236.417& 1.58E-03&      &      \\
 0&  1& 4237.563& 2.73E-03& -33.4&  21.2\\
 0&  2& 4239.381& 2.73E-03& -28.1&  18.7\\
 0&  3& 4242.116& 2.73E-03& -21.0&  29.6\\
 1& 10& 4243.787& 1.56E-03&      &      \\
 0&  4& 4245.779& 2.73E-03& -12.9&   9.5\\
-1&  2& 4247.573& 5.47E-04&  -8.9&      \\
 0&  5& 4250.387& 2.74E-03&  -6.0&      \\
 1& 11& 4250.709& 1.55E-03&      &      \\
-1&  3& 4254.389& 7.83E-04& -12.8&      \\
 0&  6& 4255.959& 2.74E-03&  -5.0&      \\
 1& 12& 4258.853& 1.54E-03&      &      \\
-1&  4& 4262.121& 9.15E-04&  -9.4&      \\
 0&  7& 4262.519& 2.74E-03&      &      \\
 1& 13& 4268.281& 1.53E-03&      &      \\
 0&  8& 4270.096& 2.75E-03&      &      \\
-1&  5& 4270.780& 1.00E-03& -10.3&      \\
 0&  9& 4278.724& 2.75E-03&      &      \\
 1& 14& 4279.063& 1.52E-03&      &      \\
-1&  6& 4280.385& 1.06E-03&      &      \\
 0& 10& 4288.440& 2.76E-03&      &      \\
-1&  7& 4290.948& 1.11E-03&      &      \\
 0& 11& 4299.286& 2.77E-03&      &      \\
-1&  8& 4302.502& 1.14E-03&      &      \\
-1&  9& 4315.078& 1.17E-03&      &      \\
  &   &         &         &      &      \\
\hline
\hline
\multicolumn{6}{l}{$f(J',J'')=f_{abs}$ is the absorption oscillator strength} \\
\multicolumn{6}{l}{for emission $g_{J'} f_{em} = g_{J''} f_{abs}$} \\
\end{tabular}

\end{document}